\documentclass[11pt,aps,pre,reprint,onecolumn,superscriptaddress]{revtex4-2}

\usepackage{graphicx}
\usepackage[utf8]{inputenc}  % Use utf8 instead of utf8x
\usepackage[english]{babel}
\usepackage{hyperref}
\usepackage{xcolor}  % Remove duplicate xcolor declarations
\usepackage{amsmath,amssymb,amsfonts}
\usepackage{bm}
\usepackage{booktabs}
\usepackage{multirow}
\usepackage{float}
\usepackage{algorithm}
\usepackage{algorithmicx}
\usepackage{algpseudocode}
\usepackage{listings}
\usepackage{lineno}

\usepackage{silence}
\usepackage{colortbl}

\begin{document}
\preprint{Submitted to Physical Review Fluids}
\title{Experimental analysis of the role of base blowing geometry on three-dimensional blunt body wakes}

\author{J. M. Camacho-S\'anchez}
\affiliation{Departamento de Ingenier\'{\i}a Mec\'anica y Minera, Universidad de Ja\'en, Jaén, Spain}
\affiliation{Andalusian Institute for Earth System Research, Universities of Granada, Jaén and Córdoba, Spain.}

\author{M. Lorite-D\'iez}
\email{mldiez@ugr.es}
\affiliation{Andalusian Institute for Earth System Research, Universities of Granada, Jaén and Córdoba, Spain.}
\affiliation{Departamento de Mec\'anica de Estructuras e Ingenier\'ia Hidr\'aulica, Universidad de Granada, Spain.}

\author{J.I. Jim\'enez-Gonz\'alez}
\affiliation{Departamento de Ingenier\'{\i}a Mec\'anica y Minera, Universidad de Ja\'en, Jaén, Spain}
\affiliation{Andalusian Institute for Earth System Research, Universities of Granada, Jaén and Córdoba, Spain.}

\author{C. Mart\'inez-Baz\'an}
\affiliation{Andalusian Institute for Earth System Research, Universities of Granada, Jaén and Córdoba, Spain.}
\affiliation{Departamento de Mec\'anica de Estructuras e Ingenier\'ia Hidr\'aulica, Universidad de Granada, Spain.}

\date{\today}

\begin{abstract}
This experimental study aims to investigate the effect of different base blowing configurations on the aerodynamics of a squareback Ahmed body of height $h$, width $w$ and aspect ratio $w/h>1$. Four different slot configurations through which air is injected were studied. Each configuration had the same blowing area, equivalent to 10$\%$ of the base area of the body, and was designated according to its geometric shape: square (S), vertical (V), cross (C) and horizontal (H). The same range of injected flow rates, $C_{q}$, was tested for each slot geometry at a Reynolds number $Re$= 65000, with corresponding wind tunnel measurements of aerodynamic forces, base pressure and wake characteristics using Particle Image Velocimetry (PIV). Our experiments revealed the effect of blowing on the near wake and consequently on the base pressure and drag of the Ahmed body. In particular, the geometry of the slots was shown to be a crucial factor in influencing aerodynamics, especially at blowing flow rates close to $C_{q,opt}$, which is the blowing flow rate that provides the minimum drag coefficient. The centered square slot (S) is the configuration that achieves a greater drag reduction. This behavior is attributed to an elongation in the recirculation region behind the Ahmed body, a reduction in the backflow inside the recirculation bubble, and a decrease in the wake asymmetry associated with the Reflexional Symmetry Breaking (RSB) mode. Conversely, the vertically oriented slot geometries, such as the cross (C) and the vertical (V) configurations, showed limited drag reduction capability while maintaining or even intensifying the wake asymmetry. The horizontal (H) slot represented an intermediate case, mitigating the wake asymmetry to a large extent, but proving to be less effective in reducing the drag than the square case. The hierarchy of the blowing configurations was dictated by the modifications induced in the near wake behind the Ahmed body and the distance from the blowing injection to the shear layers enclosing the recirculation region, which influenced the asymmetry of the wake and the filling/emptying of the recirculation region, respectively.
\end{abstract}
\maketitle
%%%%%%%%%%%%%%%%%%%%%%%%%%%%%%%%%%%%%%%%%%%%%%%%%%%%%%%%%
%%%%%%%%%%%%%%%%%%%%%%%%%%%%%%%%%%%%%%%%%%%%%%%%%%%%%%%%%

\section{Introduction}\label{sec:Intro}
%\linenumbers
%% MOTIVATION
The goal of reducing the drag of simplified blunt bodies is motivated by the large environmental impact caused by the road transportation industry, which represents 30$\%$ of the greenhouse gas emissions in the EU \citep{EEA21}. Heavy vehicles, whose main flow characteristics can be studied using the simple Ahmed body model \cite{Ahmed1984, Choi2014}, are responsible for a quarter of that figure \citep{EEA21}. Due to their specific practical requirements, they exhibit poor aerodynamic performance. In fact, they spend up to 70$\%$ of their energy consumed to overcome aerodynamic drag \citep{NRC2010}, and approximately 25$\%$ of this drag comes from the rear of the vehicle \citep{Wood03}. The blunt shape of trucks leads to massive flow separation at their trailing edges, generating turbulent three-dimensional wakes characterized by complex recirculating regions and low-pressure zones, which determine the aerodynamic drag. 

%% SEMINAL BLOWING WORKS (2D AND AXY)
The injection of a continuous flow rate into the base of a bluff body, commonly known as base blowing or base bleed, has proven to be an effective mechanism for reducing aerodynamic drag. The seminal works of \citet{Wood64} and \citet{Bearman67} investigated base blowing as a drag reduction strategy for two-dimensional (2D) blunt bodies, achieving significant base pressure recoveries by mitigating the unsteadiness of vortex shedding in the wake. 
In addition, the blown jet increases the formation length of the emitted vortices until it reaches a critical flow rate, where the high momentum of the jet decreases the base pressure of the body, increasing the aerodynamic drag.  Following this idea, a homogeneous base bleed was applied to axisymmetric blunt bodies in \cite{Sevilla04,Sanmiguel2009,Bohorquez2011} by combining experiments, numerical simulations, and stability analysis. These studies showed that the base bleed was also able to stabilize the unsteady vortex shedding behind axisymmetric bodies by decreasing the backflow velocity in the near wake.

%% 3D WAKES
However, the three-dimensional (3D) dynamics of blunt-body wakes encompass more than just vortex shedding phenomena. The reflectional-symmetry-breaking (RSB) mode, reminiscent of the first steady symmetry-breaking bifurcation at low Reynolds numbers \cite{Fabre2008, Grandemange2012,Bohorquez2011}, has emerged as a dominant flow feature influencing many drag reduction strategies \cite{Evrard2016,Haffner2021}. Under turbulent and appropriate flow conditions, this RSB mode exhibits a random long-term switching pattern between different wake-deflected states \cite{Grandemange2013a}. Particularly, in the case of simple 3D blunt bodies, such as the square-back Ahmed body, the RSB mode induces quasi-static switching between positive (P) and negative (N) states, associated with positive or negative horizontal base pressure gradients, respectively. The instantaneous asymmetry introduced by this mode substantially increases the lateral aerodynamic loads and drag of the 3D blunt bodies by approximately 10$\%$ \cite{Bonnavion2018, Haffner2020}. 

%% BASE BLOWING and 3D Wakes 
In this context, \citet{Lorite20} explored various perimetric blowing configurations to investigate the relationship between the blowing flow rates, wake entrainment, and recirculation region length for the model used in the present work. They reported two distinct ways of interaction between the base blowing and the near wake. First, a general mass regime related to the increase in the size of the recirculation bubble was proposed to explain the drag reduction observed when the blowing flow rate was small, acting as a passive scalar. Moreover, a configuration-dependent momentum regime consisting of deflation of the recirculation region leading to a base pressure decrease was also observed. Perimetric base blowing was observed to force shear layers in the latter regime, facilitating the emptying of the recirculation region. The critical blowing flow rate separating both regimes depends on the blowing slit position and near-wake asymmetry. The transition between the mass and momentum regimes was deeply investigated by employing different blowing gases in \citet{Lorite2020b}, where a blowing scaling was found. In addition, regardless of the gas density, the mechanisms of drag reduction and lengthening of the recirculation region produced by blowing in the mass regime were the same. 

%% SWEEPING JETS, BASE SUCTION, CENTRAL BASE BLOWING, SCALING
Based on these findings, recent studies have explored alternative devices, such as base suction \cite{Hsu2021} and sweeping jets \cite{Veerasamy2022}, aimed at suppressing the wake asymmetry associated with the RSB mode to reduce the drag of the square-back Ahmed body. In fact, \citet{Hsu2021} was able to symmetrize the wake by suction; however, the negative pressure induced at the base led to an increase in drag, linked to a reduction in the length of the recirculation region. The different slots tested by \citet{Veerasamy2022} showed that placing the blowing actuator in the center of the Ahmed body was optimal for reducing the aerodynamic drag. Based on this result, \citet{Khan2022} introduced a small continuous jet into the 3D wake behind the same model through several small blowing slots. In this case, blowing effectively reduced the drag by recovering the base pressure and symmetrizing the RSB mode in the mass regime. Furthermore, a proper scaling based on a limit in the blown jet momentum was proposed to collapse the critical blowing flow rates for the three tested slots. Very recently,  \citet{Khan2024} investigated the balance of fluxes entering and exiting the recirculation bubble using Stacked Stereoscopic Particle Image Velocimetry (SSPIV) across multiple planes on an Ahmed body with a central square slot, similar to that examined in \citet{Khan2022}. In their study, the authors reformulated the two regimes described by \citet{Lorite20}, introducing three distinct phases: the mass regime, favorable momentum regime, and momentum regime. The favorable momentum regime is characterized by an increase in base pressure and a reduction in drag, similar to the mass regime in~\citet{Lorite20}, with the key distinction that the momentum of the jet appears to significantly counteract backflow reentry.

The present study aims to investigate the influence of blowing-slot geometry on the three regimes in an attempt to explain the differences observed between the perimetric and central blowing slots. To this end, we explore symmetric blowing configurations with the same blowing sections to elucidate their role in statistically symmetric 3D wakes as an alternative to the asymmetric unsteady forcing of asymmetric 3D wakes studied by \citet{Haffner2021}. Finally, we analyze the performance of the different injection geometries to identify the best blowing configuration in terms of drag reduction in 3D simple blunt bodies and to determine the underlying physical mechanisms.

%% PAPER ORGANIZATION  
The remainder of this paper is organized as follows: Sect. \ref{sec:set-up} details the experimental set-up while Sect. \ref{sec:results} discusses the main experimental results, including base flow characterization (Sect. \ref{subsec:Baseline}), the global effects of base blowing (Sect. \ref{subsec:Global}) and near wake modifications induced by the air injection under the different tested configurations (Sect. \ref{subsec:NearWake}) . 
 The relationship between the different blowing configurations is discussed in Sect. \ref{sec:Discussion}, and, Sect. \ref{sec:Conclusions} summarizes the main conclusions of the work.

\section{Experimental description}\label{sec:set-up}
\subsection{Wind tunnel set-up and blowing system}
\begin{figure}[ht]
\centering
\includegraphics[width=0.94\textwidth]{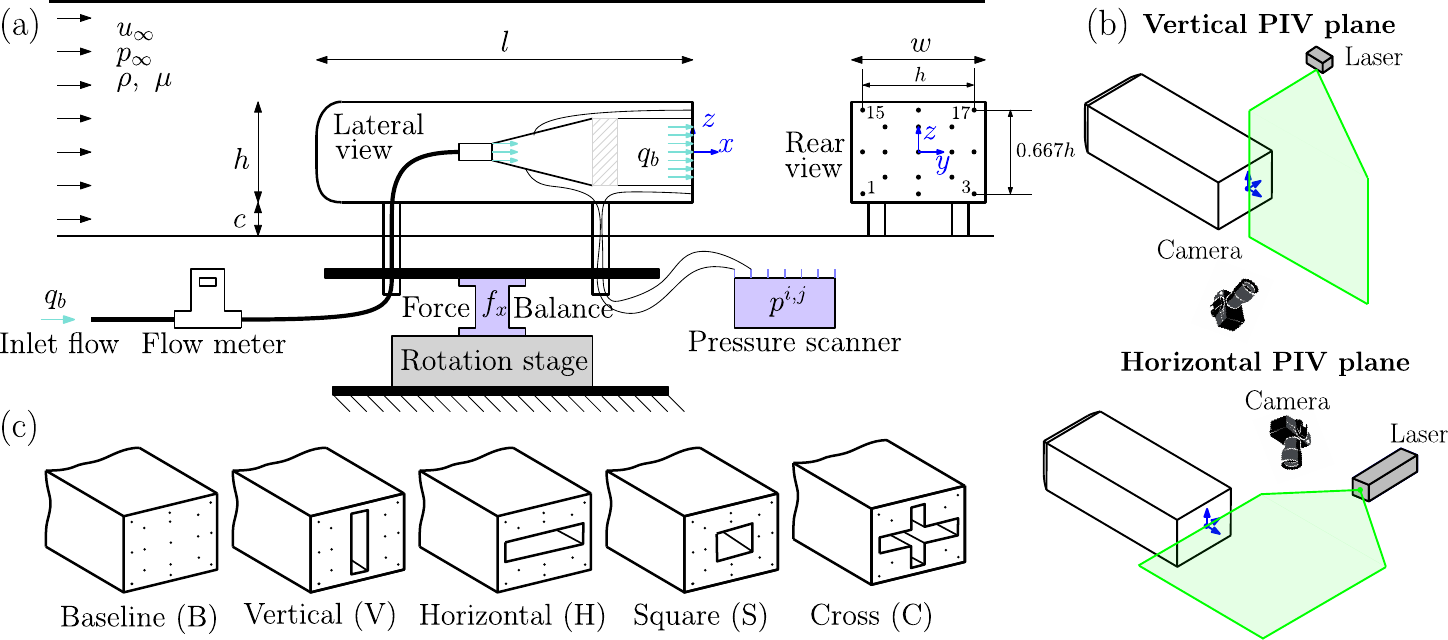}
\caption{\label{fig:setup}Sketch of the experimental set-up. (a) Lateral and rear view of the Ahmed body inside the wind tunnel. (b) Illustration of the PIV measurements in vertical and horizontal planes. (c) Set of the different base blowing configurations considered in the present work, showing the available pressure taps in each case illustrated with black dots.}
\end{figure}
We performed an experimental study on a square-back Ahmed body, as depicted in Fig.~\ref{fig:setup}, with dimensions $h=72$ mm (height), $w=97.25$ mm (width), and $l=261$ mm (length). The experiments were conducted in a recirculating wind tunnel with a nozzle with a contraction ratio of 8:1, connected to a 400 $\times$ 500 mm$^2$ test section and a blockage ratio lower than 5\%, considering the frontal area of the Ahmed body. The freestream velocity in the test section can be varied from 5 to 20 m/s, with a turbulent intensity below $1\%$. The flow uniformity along the test section was always greater than 98$\%$ for the selected freestream velocity. The body was positioned $c=20$ mm from the wind tunnel ground using four cylindrical steel tubes with an external diameter of 8 mm. A coordinate system with its origin at the center of the Ahmed body base was established, with $x$ denoting the streamwise direction, $z$ the vertical direction, and $y$ completing the right-hand trihedron. The Ahmed body was mounted on an automatic rotation stage (Standa 8SMC5-USB) with an accuracy of 0.01$^\circ$. All tests were conducted under aligned flow conditions with zero yaw angle. To align the body with the incoming flow, we sought a symmetrical distribution of the Reflectional-Symmetry-Breaking (RSB) mode for the baseline case in the horizontal direction, compensating for any asymmetry in our experimental set-up \cite{Evrard2016}. The incoming freestream velocity was established as $u_\infty=13.6$ m/s, providing a Reynolds number of $Re=\rho h u_\infty/\mu \simeq 65000$, where $\rho$ and $\mu$ are the air density and viscosity, respectively. The temperature during experiments was maintained at $T=21\pm1$ºC.

The blowing system comprised an air compressor capable of delivering a constant pressure of 9 bar up to a flow rate of 200 L/min, and a pressure regulator to control the pressure in the air feeding line. Downstream, a fine precision valve controlled and fixed the insufflated flow rate, denoted as $q_b$. To measure $q_b$, an Aalborg digital flow meter (model \#GFM57) was used with a measuring range of 0-180 L/min, accuracy within $\pm 1.5\%$ and  repeatability around $\pm 0.25\%$ of the full-scale range.  A flexible tube was introduced through one of the supports of the Ahmed body to inject the blowing air, which passed through a 140 mm long divergent-convergent nozzle with a 10 mm thick foam to ensure flow uniformity at the blowing slot (see Fig. ~\ref{fig:setup} a). The blowing flow rate coefficient $C_{q}$ is defined as $C_{q}=q_{b}/(u_{\infty} wh)$. Nineteen different flow rates were tested for each blowing configuration in the range $C_q \in [0, 0.0315]$, and the measurements were repeated three times for each flow rate to ensure the robustness of the results. The data reported here are the averages of the three tests performed under each experimental condition.

We tested four different symmetric blowing slots geometries located at the center of the Ahmed body base: horizontal (H), vertical (V), cross (C), and squared (S), as illustrated in Fig.~\ref{fig:setup}(c). The blowing area, $S_b$, was the same in all configurations, with $S_b/hw=0.1$, which allowed the exploration of different spatial distributions of the base blowing.

\subsection{Forces, pressure and velocity measurements}\label{subsec:Measurements}

The aerodynamic drag force of the Ahmed body in the streamwise direction, $f_x$, was measured using a 6-axial force balance (model SRI M3703A) with a 50 N range in $x$ and $y$ directions, and 100 N range in $z$ direction. The force balance can operate with a crosstalk lower than 2$\%$ and a non-linearity smaller than 0.5$\%$ of the sensor’s full range. The load cell was connected to the Ahmed body through four cylindrical supports that aligned the force balance directions with our coordinate system (see Fig.~\ref{fig:setup}). Force measurements were performed during 120 s at an acquisition rate of 200 Hz. The instantaneous non-dimensional drag coefficient is defined as

\begin{equation}
c_x=\frac{2(f_x-f_{x,b})}{\rho u_\infty^2 wh},
\end{equation}
where $f_{x,b}$ represents the thrust force produced by flow injection at the base. The resulting force coefficient $c_x$ obtained by subtracting $f_{x,b}$ from the global measured forces in the experiment, allows the quantification of the wake-related aerodynamic improvement given by blowing-driven changes. In addition, the computed averaged values of the drag have an associated uncertainty below $\pm 0.002$. 

The base pressure on the Ahmed body was measured using a 64-channel pressure scanner with a 4" H$_{2}$O range, connected up to 17 pressure taps, $p^{i,j}$, distributed in the body base, as in the rear view of the model depicted in Fig.~\ref{fig:setup}(a). The pressure measurements were acquired at a frequency of 200 Hz, with each tap providing an accuracy of 0.6 Pa. The pressure was non-dimensionalized using pressure coefficients, defined as

\begin{equation}
c_{p}^{i,j}=\frac{2(p^{i,j}-p_{\infty})}{\rho u_\infty^2},
\end{equation}
From these values, the base drag coefficient \citep{Roshko1993} was estimated by means of 
\begin{equation}
c_{B}=-\frac{1}{n_a} \sum_{i=1}^{n_a} c_{p}^{i,j}
\end{equation}
where $n_a$ represents the number of available taps (note that depending on the blowing geometry, certain tap locations may not be accessible, varying the measurement points for computing the base pressure). The uncertainty associated with $c_{p}^{i,j}$ values was less than $\pm0.001$ in the present experiments. Additionally, the base pressure gradients were calculated as
\begin{equation} \label{eq:gy}
g_y=h \frac{\partial c_p}{\partial y} \simeq \frac{h}{n_i-1}\sum_{j=1}^{n_i-1}\left[\frac{\frac{1}{n_j}\sum_{i=1}^{n_{j}}c_p^{i,j}-\frac{1}{n_{j+1}}\sum_{i=1}^{n_{j+1}} c_p^{i,j+1}}{(y_j - y_{j+1})}\right],
\end{equation}
\begin{equation} \label{eq:gz}
g_z=h \frac{\partial c_p}{\partial z} \simeq \frac{h}{n_j-1}\sum_{i=1}^{n_j-1}\left[\frac{\frac{1}{n_i}\sum_{j=1}^{n_{i}}c_p^{i,j}-\frac{1}{n_{i+1}}\sum_{j=1}^{n_{i+1}} c_p^{i+1,j}}{(z_i - z_{i+1})}\right],
\end{equation}
Here, $g_y$ (respectively $g_z$) represents the dimensionless horizontal (vertical) base pressure gradient, $c_p^{i,j}$ denotes the pressure coefficient at the tap located at row $i$ and column $j$, with $i=1$ the topmost row ($z$ positive) and $j=1$ the rightmost column ($y$ positive), and $h$ is the height of the Ahmed body. Furthermore, $n_{i}$ indicates the total number of taps in row $i$ and $n_j$ is the number of pressure taps in column $j$, where $y_j$ and $z_i$ are the $y$-coordinates of column $j$ and the $z$-coordinates of row $i$ respectively. Equation \eqref{eq:gy} (respectively Eq.~\ref{eq:gz}) represents the mean horizontal (respectively vertical) gradient of the average values in each column (respectively row). Despite the potential limitations in tap accessibility, the selected tap arrangements provide relevant and representative measurements for accurately estimating the base drag and pressure gradients. This assumption holds since the wall pressure distribution in the separated area remains approximately constant in one direction at first order and affine in the perpendicular direction (direction of asymmetry) \citep[see e.g.][]{Barros2017}. Besides, the total base pressure asymmetry can be computed as 
\begin{equation}
g=\sqrt{g^2_{y}+g^2_{z}}.
\end{equation}

Furthermore, we performed experiments to characterize the velocity field $\textbf{u}=(u_{x}, u_{y}, u_{z})$ within the near wake behind the Ahmed body. In particular, we performed 2D-2C PIV measurements on a vertical plane ($y=0$) to obtain the ($u_{x}, u_{z}$) velocity components and at a horizontal plane ($z=0$) to obtain the ($u_{x}, u_{y}$) velocity components (see Fig.~\ref{fig:setup}b). For our experimental set-up, we used a high-performance double pulsed laser model Litron LPY704-100 PIV (100 mJ/pulse, 100 Hz, 532 nm) to generate a laser sheet in the wind tunnel. To precisely control the position and thickness of the laser sheet ($\simeq 1$mm) within the measurement volume, we employed an optical arm, cylindrical lens (-10 mm), and collimator. The laser and an Imager MX 25M (25 Mpx) camera, equipped with a 105 mm f/5.6 lens, were synchronized with a LaVision{\textregistered} PTU-X. Before acquiring the images, we performed a binning operation ($\times$2, $\times$2) on the camera, resulting in a final image resolution of 2560 $\times$ 2560 pixels. PIV calibration was conducted using a checkerboard calibrated plate to properly focus the camera and set the laser plane position and width, obtaining a scale of 23.31 px/mm. Moreover, the incoming wind was seeded using an oil droplet generator with tracers of diameter $\simeq 2~\mu$m. The laser sheet was pulsed with a time delay of $dt = 31~\mu$s, and the set-up acquired 500 pairs of images at 5 Hz, ensuring an adequate number of images to obtain the time-averaged velocity fields. After correctly setting the region of interest, velocity vectors were obtained from the multi-pass PIV correlation in the interrogation windows of 64 $\times$ 64 pixels with an overlap of 75$\%$. The resulting velocity vectors were spatially distributed in a grid of $192 \times 256$ points in the horizontal plane and $156 \times 223$ points in the vertical plane  with a resolution below $1\%$ of the height of the body for both measurement planes. 

Our experiments included PIV measurements at four different flow rates, $C_q=[0, 0.01, 0.021, 0.031]$ for each blowing configuration and the corresponding baseline body (B).Finally, we checked our velocity measurements based on correlation-based uncertainty \citep{Wieneke2015} and obtained errors typically below 1\% of the freestream velocity in our measurement region. The zones with the highest velocity gradient have an uncertainty of approximately 3\% for the streamwise velocity, whereas the uncertainty is around 2\%  for the other velocity components in slow velocity regions. 

Henceforth, $h$, $u_{\infty}$, and $h/u_{\infty}$ are used as the characteristic length, velocity, and time scales, respectively. We denote with an asterisk $^*$ the dimensionless variables defined from these characteristic scales. A time series of variables is denoted by the lower case letter $v$, time averages are denoted by the upper case letter $V$, and fluctuations around the time average of a given variable are denoted by $\hat{v}=v-V$. Moreover, the spatial average of a given variable will be expressed as $\langle v \rangle$. In addition, all vector variables, whether they represent the time series ($\Vec{v}=\bm{v}$) or time averages ($\Vec{V}=\bm{V}$), are shown in bold. 

Finally, conditional statistics will be applied to certain measurements to examine the characteristics of the $P$ or $N$-states of the RSB mode \cite{Grandemange2013a}. These states correspond to events which the horizontal base pressure gradient is $g_{y}>0$ ($P$-state) or $g_{y}<0$ ($N$-state). For the remainder of this paper, we use a superscript $P$ or $N$ to denote the conditional averaging of variables based on this criterion.

 %%%%%% RESULTS %%%%%%
\section{Results}\label{sec:results}
\subsection{Baseline case features}\label{subsec:Baseline}

\begin{figure}[t]
\centering
\includegraphics[width=1\textwidth]{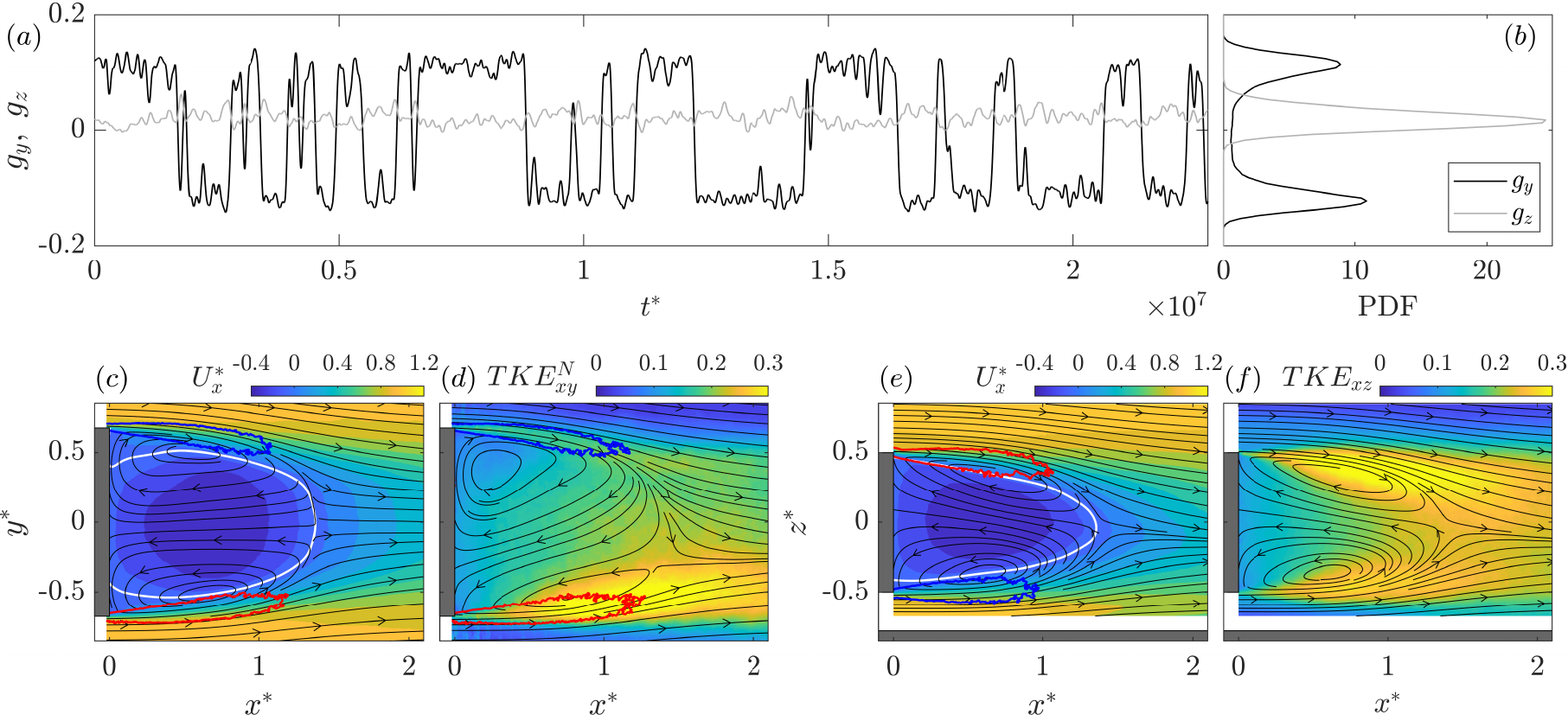}
\caption{\label{fig:PIV_Gy_Gz_Ref} Wake characterization for the baseline configuration (B): (a) temporal evolution of the horizontal, $g_y$, (black line) and vertical, $g_z$, (grey line) base pressure gradients and (b) their corresponding PDFs. (c, e) Time-averaged streamwise velocity contours, $U_x^*$, along with averaged streamlines in the horizontal plane at $z^*=0$ and in the vertical plane at $y^*=0$, respectively. (d) Conditional averaged turbulent kinetic energy ($TKE^N_{xy}$) contours and mean streamlines corresponding to the $N$-state in a horizontal plane at $z^*=0$, and (f) mean turbulent kinetic energy ($TKE_{xz}$) contours and streamlines in the vertical plane at $y^*=0$. The solid white lines indicate the recirculation region border, using the isoline, $U^{*}_{x}=0$, while red/blue zones illustrate the shear layers (iso-contours of vorticity, $\Omega^{*}_{i}=\pm 4$) in each plane.}
\end{figure}
The reference body studied in this work is a square-back Ahmed body without blowing slots at the base, i.e., $C_q=0$ (baseline configuration in Fig.~\ref{fig:setup}c). The wake characteristics of this configuration is shown in Fig.~\ref{fig:PIV_Gy_Gz_Ref}, which illustrates the temporal evolution of the horizontal, $g_y$, and vertical, $g_z$, base pressure gradients (Fig.~\ref{fig:PIV_Gy_Gz_Ref}a), along with their corresponding Probability Density Functions, PDFs  (Fig.~\ref{fig:PIV_Gy_Gz_Ref}b). Our measurements reveal the presence of the Reflectional-Symmetry-Breaking (RSB) mode in the horizontal direction, characterized by a bistable behavior between two states: positive ($P$) when $g_y>0$ and negative ($N$) when $g_y<0$. The corresponding PDF displays two equally probable peaks of $g_y$, which indicates the  experimental symmetry of the set-up \cite{Evrard2016}. Consequently, the near wake depicts a symmetric mean recirculation region in the horizontal plane, as illustrated in Fig.~\ref{fig:PIV_Gy_Gz_Ref}(c) by the contours of the mean streamwise velocity $U_x^*$ and the mean streamlines in a horizontal plane at $z^*=0$. By analyzing the random switches between the $N$ and $P$-states, it is possible to calculate the conditional average of each state for the PIV measurements. %We compute the spatial average of $\langle u_{y} \rangle$ within the recirculation region for each instant. Subsequently, instants associated with a positive (respectively negative) spatial average of $u_y$ correspond to the $P$ (respectively $N$) wake deflected state.
In that regard, Fig.~\ref{fig:PIV_Gy_Gz_Ref}(d) shows the near wake corresponding to the $N$-state of the RSB mode with the corresponding flow streamlines. It also includes the spatial distribution of the velocity fluctuations, represented by the conditional averaged turbulent kinetic energy, defined here as $TKE^N_{xy}=\sqrt{\hat{U}^{*2}_{x}+\hat{U}_{y}^{*2}}$ corresponding to the $N$-state in the $x^*-y^*$ plane.

When the horizontally deflected $N$-state is present, the wake exhibits significant asymmetry, with two recirculation cores of different shapes located at different streamwise positions. In other words, a wide recirculation region is fixed on the $y^{*}>0$ side of the base, which induces a pressure decrease in this region, resulting in a negative horizontal base pressure gradient. Conversely, the velocity fluctuations are concentrated on the opposite side of the wake, $y^{*}<0$, as vortex emission occurs in this region (see the three-dimensional representation of the wake behind the squareback Ahmed body from \citet{Pavia20} and \citet{Khan2024}). For simplicity, hereafter, we will omit the phase averaging of the $P$-state, which mirrors the $N$-state. 

We also analyzed the shear layers enclosing the recirculation region. To that aim, Figs. \ref{fig:PIV_Gy_Gz_Ref}(c-d) depict, in red (bottom) and in blue (top), isolines of constant mean vorticity calculated as $\Omega_{z} = {dU_{y}}/{dx}-{dU_{x}}/{dy}$ (and made dimensionless as $\Omega^{*}_{z}=\Omega_{z}h/u_{\infty}$). The shear layers (depicted by vorticity isocontours $\Omega^{*}_{z}=\pm 4$) and velocity fluctuations govern the entrainment of momentum in the recirculation region, affecting the pressure balance in the near wake \cite{Khan2024}. Under the $N$-state (Fig. \ref{fig:PIV_Gy_Gz_Ref}d), the asymmetry of the wake leads to a curved shear layer on the $y^{*}>0$ side of the base, generating the aforementioned large recirculation zone. This wake asymmetry reduces the curvature of the opposite shear layer on the $y^{*}<0$ side, resulting in additional wake excitation and a thicker shear layer characterized by significant flow fluctuations. Moreover, the recirculation bubble limits are indicated by white solid lines in Figs. \ref{fig:PIV_Gy_Gz_Ref}(c-f). These limits, defined as the isocontour of streamwise velocity $U^{*}_{x}=0$, allow the determination of the spatial extent of the recirculation bubble and derived parameters such as the recirculation length, $L_r^*$, given by the distance from the body base, $x^*=0$, and the farthest downstream point belonging to the recirculation region border, $U^{*}_{x}=0$.

On the other hand, the time evolution of the vertical base pressure gradient $g_z(t^*)$ remains nearly constant, showing a slightly positive mean value induced by the wind tunnel floor (see Fig.~\ref{fig:PIV_Gy_Gz_Ref}b). The slight asymmetry in the near wake is observed in Fig.~\ref{fig:PIV_Gy_Gz_Ref}(e), where the lower recirculation zone stays closer to the base than the upper one, thus resulting in a value of $g_z>0$. The shear layers are also illustrated in this plane by isolines of constant vorticity $\Omega^{*}_y= \Omega_{y}h/u_{\infty}=\pm 4$, where $\Omega_{y} = {dU_{z}}/{dx}-{dU_{x}}/{dz}$. Despite the slight asymmetry induced by the ground, the magnitude and extent of the flow fluctuations around the recirculating region in the vertical plane are similar on both sides of the $z^*$-axis, as illustrated in Fig.~\ref{fig:PIV_Gy_Gz_Ref}(f).

\begin{table}[t]
\centering
\begin{tabular}{cc|cc|cc}
\hline
\multicolumn{2}{c|}{Drag} & \multicolumn{2}{c|}{RSB mode} & \multicolumn{2}{c}{PIV} \\ \hline
$C_x$        & 0.412 $\pm$ 0.002       & $G^N_{y}$         & 0.120  $\pm$ 0.001     & $L_r^*$      & 1.36 $\pm$ 0.01     \\
$C_B$         & 0.189 $\pm$ 0.001            & $G_z$         & 0.025  $\pm$ 0.001        & - \\ \hline
%$\langle TKE^{xy}_{N,SL} \rangle$      & 0.284 $\pm$ 0.001    \\ \hline
%-        & -        & -       & -       & $\langle TKE^{xz}_{SL}\rangle$      & 0.277 $\pm$  0.002     \\ \hline
\end{tabular}
\caption{{Reference averaged values from the baseline configuration: drag coefficient $C_x$, base drag coefficient $C_B$, base pressure gradients $G^N_y,G_z$ and recirculation length $L^{*}_r$.}}
\label{tab:RefData}
\end{table}

The different time-averaged values of the main parameters are computed and summarized in Table \ref{tab:RefData} for the baseline configuration $C_q=0$, which includes the drag coefficient $C_x$, base drag coefficient $C_B$, base pressure gradients $G^N_y$ and $G_z$, and recirculation length $L^*_r$. As the baseline case presents equiprobable horizontal bistability, the mean horizontal pressure gradient $G_y$ is zero; therefore, the value given in Table \ref{tab:RefData} corresponds to the conditional average for the $N$-state, $G^N_{y}$. In addition, the recirculating length $L^{*}_r$ is obtained for the vertical plane to avoid the additional uncertainty introduced by the RSB bistability in the horizontal plane. Despite slight experimental variations arising from differences in the ground clearances or cylindrical supports, our values are similar to those reported in the literature \cite{Grandemange2013a, Lorite2020b, Veerasamy2022, Khan2022}. In the following sections, we will analyze the changes induced by the different blowing configurations on the wake topology, and consequently, on the aerodynamic forces and pressure distribution.

\subsection{Global blowing effects}\label{subsec:Global}
%Drag reduction, base pressure gradient, Snapshots PIV
\begin{figure}[t]
\centering
\includegraphics[width=1\textwidth]{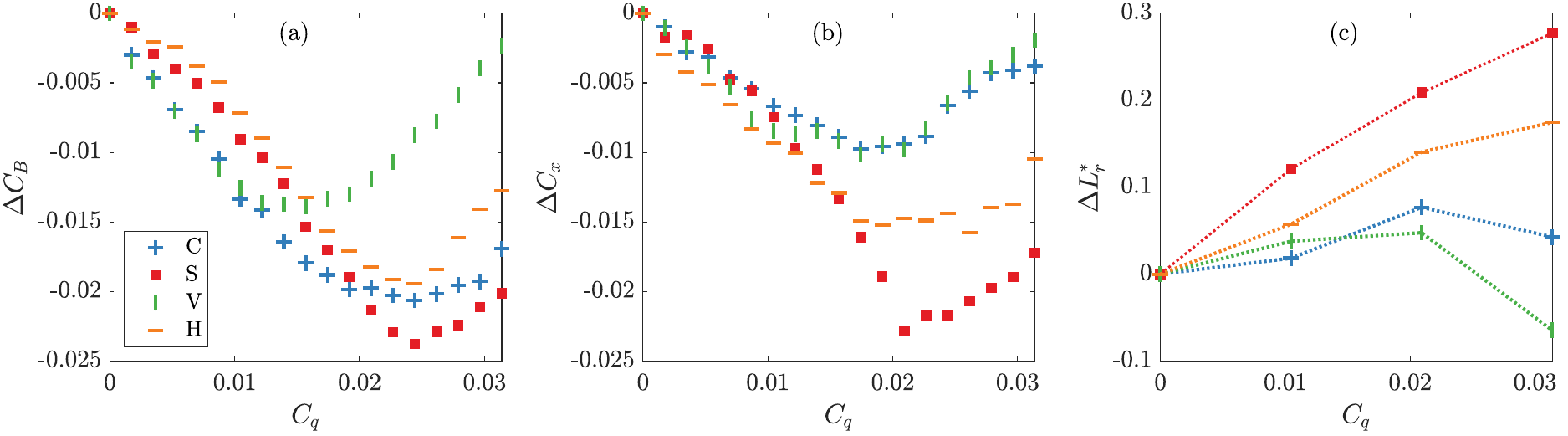}
\caption{\label{fig:ACx_ACB} Evolution with $C_q$ of relative variations of (a) the base drag coefficient $\Delta C_B$, (b) the drag coefficient $\Delta C_x$ and (c) the recirculation bubble length $\Delta L_r^*$.}
\end{figure}
%% Presentation of the figure
We will first analyze the influence of different blowing configurations on the main aerodynamic features of the wake. To that aim, Fig.~\ref{fig:ACx_ACB} shows, for each blowing arrangement, the evolution with the blowing flow rate coefficient $C_q$ of the relative variations of the base drag coefficient, $\Delta C_B(C_q)$, the drag coefficient, $\Delta C_x(C_q)$, and the recirculation bubble length, $\Delta L_r^*(C_q)$. These variations are computed as $\Delta C_B^{k}(C_q) =C_B^{k}(C_q)-C_B^k(C_q=0)$, $\Delta C_x^{k}(C_q) =C_x^{k}(C_q)-C_x^k(C_q=0)$ and $\Delta L_r^{*k}(C_q)=L_r^{*k}(C_q)-L_r^{*k}(C_q=0)$, denoting $k=$ [S, V, H, C] the different blowing arrangement. It should be noted that open passive slits with no blowing rate (i.e., $C_q=0$) had no appreciable effect on the results, and yielded values similar to the baseline case.

%% Mass regime
The analysis of Fig.~\ref{fig:ACx_ACB} provides a valuable insight into the aerodynamic modifications induced by the increasing blowing for the different arrangements. Notice that, all configurations exhibit similar behaviour, whereby the drag  initially decreases with increasing blowing rate, as the base pressure recovers, until a minimum value is reached, from which this trend subsequently reverses and the drag begins to increase with $C_q$. Therefore, two different regimes can be identified for increasing values of $C_q$, namely, the \textit{mass} and \textit{momentum} regimes, as denoted by \citet{Lorite20}. The mass regime is characterized by the injection of a continuous jet with negligible momentum, which passively fills the recirculation region. This translates into lengthening of the recirculating bubble as $C_q$ increases, as illustrated by the evolution of $\Delta L_r^{*k}$ shown in Fig.~\ref{fig:ACx_ACB}(c). Consequently, the drag and base drag coefficients decrease monotonically with $C_q$. Within this regime, the evolutions of $\Delta C_x(C_q)$ and $\Delta C_B(C_q)$ nearly collapse for all configurations, with no effect of the blowing geometry on drag reduction. 
%% Favourable momentum regime
However, the critical blowing flow rate $C^{k}_{q,opt}$ for which the maximum drag reduction is achieved, that is, $\Delta C^{k}_{x,max}$ and $\Delta C^{k}_{B,max}$ depends on the blowing configuration. 
The \textit{favorable momentum} regime introduced by \citet{Khan2024} is not evident in our measurements, where the decrease in the drag and base drag coefficients is approximately monotonous without any inflection point.

According to our measurements, the momentum of the blowing jet becomes relevant for values of $C_{q}$ close to $C^{k}_{q,opt}$, where clear differences arise between the tested blowing configurations, leading to different values of $C^{k}_{q,opt}$ and, consequently, of  $\Delta C^{k}_{x,max}$ and $\Delta C^{k}_{B,max}$. In particular, blowing through the squared slit (S) stands out as the most efficient system, as it extends the mass regime towards higher values of $C_q$, thus providing the largest reductions in the present study, i.e. approximately $5\%$  for $C_x$ and 12$\%$ for $C_B$, at $C_{q} \simeq 0.0244$. This favorable injection regime also expands the recirculation region, as shown in Fig. ~\ref{fig:ACx_ACB}(c). In fact, the hierarchy between blowing configurations is similar for both representations $\Delta L^{*}_r(C_q)$ and $\Delta C_x(C_q)$, suggesting that the drag reduction mechanism is mostly governed by the elongation of the recirculation region and the consequent base pressure changes (the increase in the recirculation region also modifies the flow exchange between the recirculation region and freestream flow, as will be discussed later).

%% Momentum regime
After the minimum drag is achieved at $C_{q,opt}^k$, the momentum regime begins. As detailed in the literature, in this regime, the momentum of the jet is sufficiently large to be unstable, strongly modifying the flow balance of the recirculation region, which acts by increasing the associated drag.
For the highest blowing rates, the force measurements may also be affected by the nonlinear interaction of the blowing jet with the wake flow, which cannot be assessed by simply subtracting the blowing thrust explained in Sect. \ref{sec:set-up}.

\subsection{Near wake modifications by base blowing}\label{subsec:NearWake}
In this section, we first analyze, in Sect. \ref{sec:AvgWake}, the blowing effect on the averaged wake in terms of absolute and relative changes. The balance of fluxes entering and leaving the recirculation region is discussed in Sect. \ref{sec:fluxes} and the impact of the blowing on the wake fluctuations is described in Sect. \ref{sec:tke}. Finally, the influence of the blowing configuration on the RSB mode is analyzed in Sect. \ref{subsec:RSB}.

\subsubsection{Effects of blowing in the recirculating bubble} \label{sec:AvgWake}
The above comprehensive analysis will now be complemented by the analysis of the near-wake flow field to investigate the mechanisms governing the recirculation region. For this purpose, we conducted PIV measurements for selected values of the blowing coefficient $C_{q}=[0, 0.01, 0.021]$, which are representative of the mass regime, $C_{q}=0.01$, and approximately the optimal blowing flow rate, $C_q = 0.021 \simeq C^{k}_{q,opt}$. Note that in the case of configuration V, $C_{q}=0.021$ already falls within the momentum regime, as shown in Fig.~\ref{fig:ACx_ACB}. However, this will be included in the analysis for comparison and illustration purposes. In addition, the near-wake corresponding to the baseline (reference) configuration, described in Section \ref{subsec:Baseline}, is also plotted.

\begin{figure}[t]
\centering
\includegraphics[width=1\textwidth]{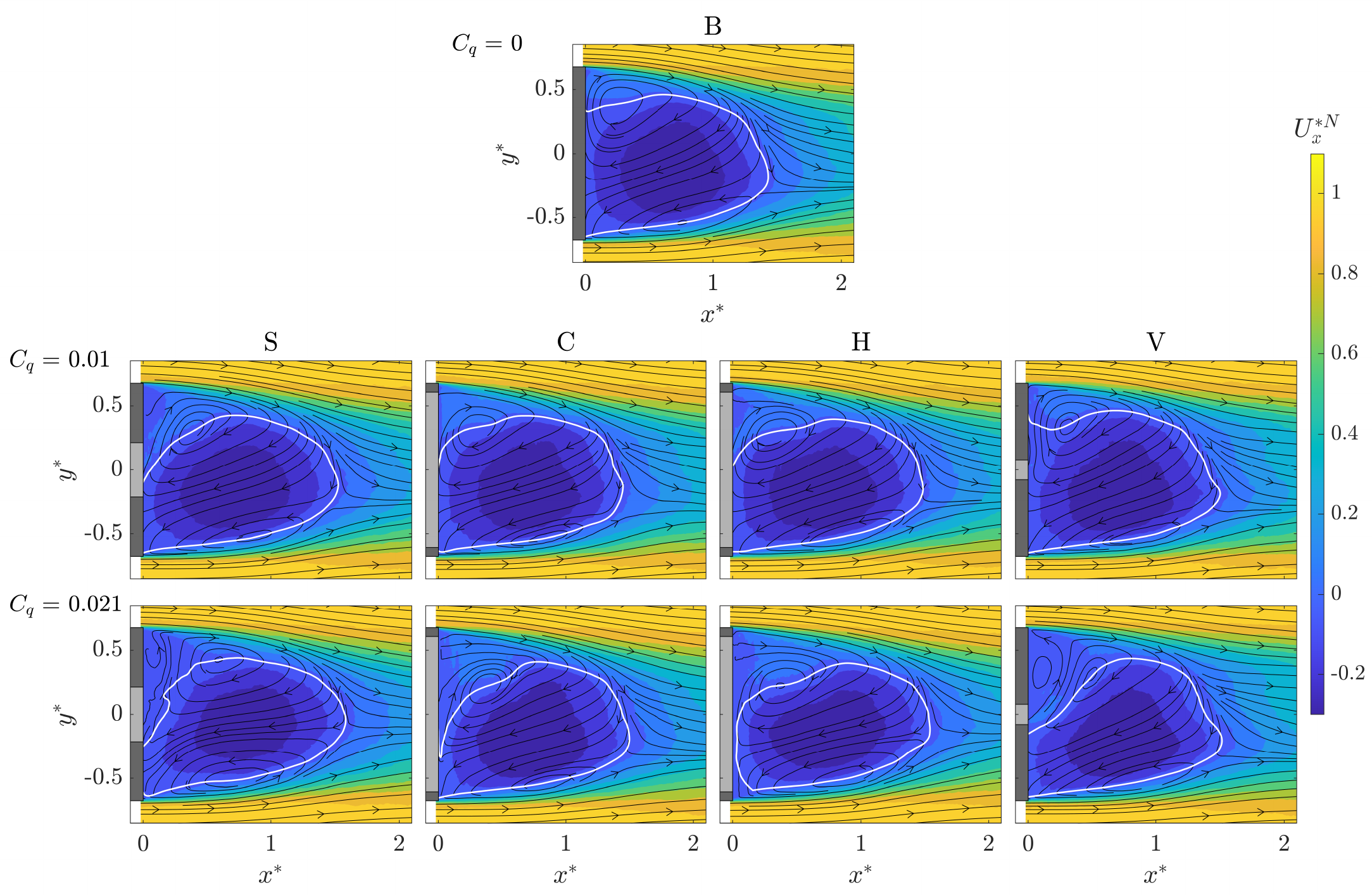}
\caption{\label{fig:HORPIV_AVG}Averaged flow field in the horizontal plane, $z^{*}=0$, corresponding to the baseline case (B) and the different blowing arrangements (S, C, H, V) for the flow rates $C_{q}=[0.01, 0.021]$ in the $N$-state. The flow is described by the contours of the conditionally averaged streamwise velocity $U^{*N}_{x}$ and flow streamlines. White isoline represents $U^{*N}_x=0$. The body base is shown in dark gray while the blowing slot is displayed in light gray.}
\end{figure}

\begin{figure}[t]
\centering
\includegraphics[width=1\textwidth]{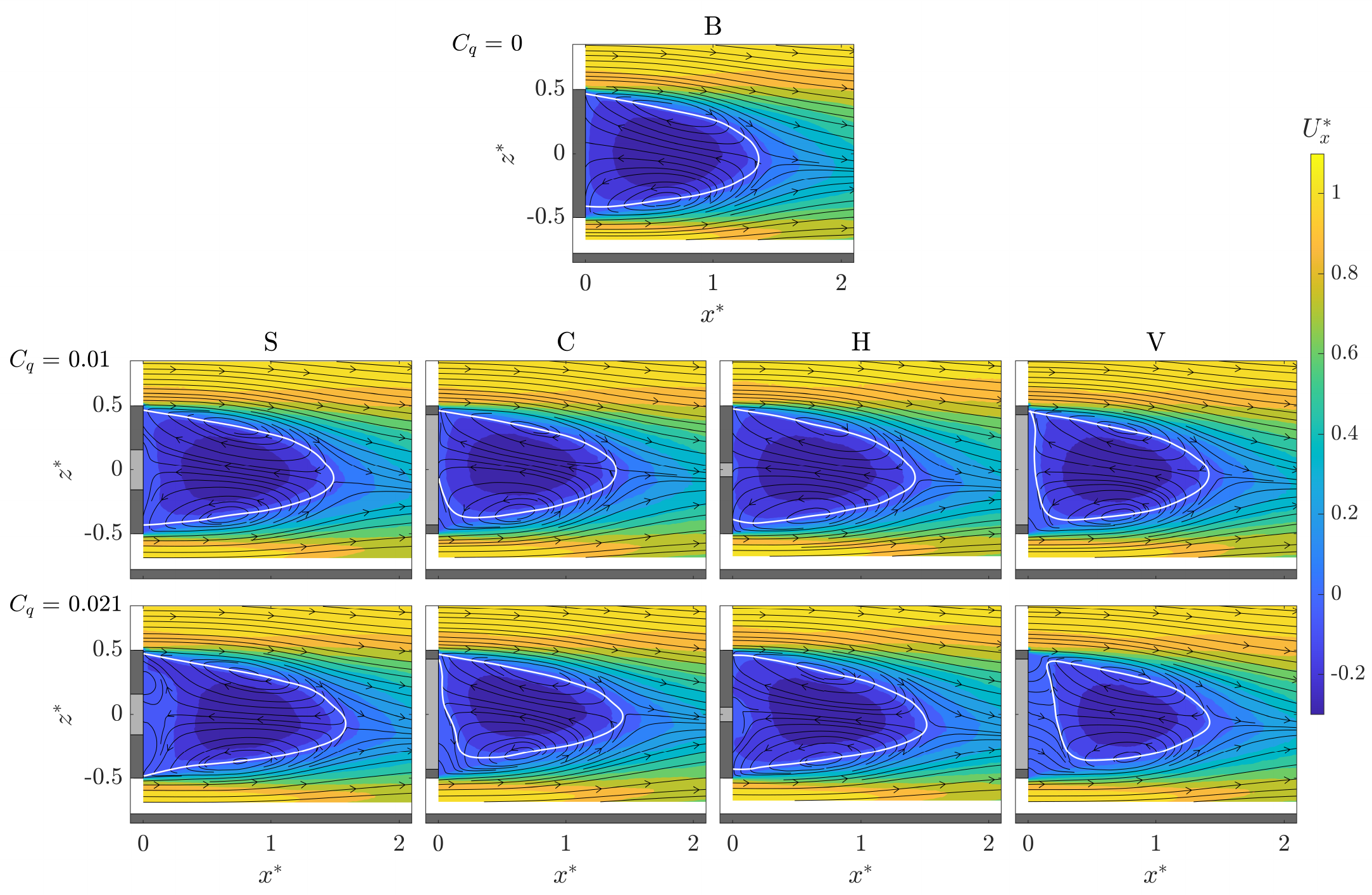}
\caption{\label{fig:VERPIV_AVG} Averaged flow field in the vertical plane, $y^{*}=0$, corresponding to the baseline case (B) and the different blowing arrangements (S, C, H, V) for the flow rates $C_{q}=[0.01, 0.021]$. The flow is described by the contours of the averaged streamwise velocity $U^{*}_{x}$ and averaged streamlines. White isoline represents $U^{*}_x=0$. The body base is shown in dark gray while the blowing slot is displayed in light gray.}
\end{figure}

%% Description of the changes in the mean flow
Changes in the near wake region induced by different blowing systems and the selected flow rates are displayed in Figs.~\ref{fig:HORPIV_AVG} and \ref{fig:VERPIV_AVG}, where the contours of the average streamwise velocity, $U^{*}_{x}$, and flow streamlines are represented in the horizontal, $z^*=0$, and vertical, $y^*=0$, planes, respectively. Furthermore, the location of the recirculating bubble is illustrated by the isoline $U^{*}_{x}=0$. 
The horizontal direction analysis is applied to the conditionally averaged flow corresponding to the $N$-state, since the global average is not representative of the preferred deflected wake, as discussed by \citet{Khan2024}.  

Let us first analyze the effect of low blowing rates, e.g. $C_{q}=0.01$, which is within the mass regime. Although the near wake is similar to that corresponding to the baseline case, it is seen that, in general, the recirculating region extends further downstream inducing base pressure recovery, as reported in Fig.~\ref{fig:ACx_ACB}(c) (the V configuration shows a distinct topology). 

When the blowing rate increases close to \(C^{k}_{q,\text{opt}}\), i.e. \(C_{q} = 0.021\), the momentum of the blowing jet becomes significant and there is a clear influence on the topology of the recirculation region. In particular, the jet can be identified near the base of the body for both the horizontal and vertical planes in Figs.~\ref{fig:HORPIV_AVG} and \ref{fig:VERPIV_AVG}. It can also be seen that the large recirculation core, located on the half-positive side $y^{*} > 0$ of the horizontal view of the $N$-state, deflects the blown jet due to the existing transverse base pressure gradient, which prevents the jet from filling the whole recirculation bubble. As a result, only a small part of the base, in the $y^*<0$ half, is exposed to the backflow, especially in the case of arrangements that concentrate the injection on the $y^*=0$ axis. When the blowing geometry is considered, some differences on the blowing effect arise. For example, in the H configuration, which acts as a horizontal plane jet, blowing separates the recirculation area from the base. It is also noted that the recirculation bubble asymmetry in the horizontal plane is slightly reduced for both H and, particularly, the S configurations. In addition, for the V configuration, the selected blowing flow rate, $C_{q}=0.021$, belongs to the momentum regime, which substantially modifies the near wake flow. As illustrated in Fig.~\ref{fig:HORPIV_AVG}, the asymmetry is increased by the interaction of the wake and blown jet; consequently, an additional recirculation cell is formed in the lower pressure area, $y^*>0$ (a detailed analysis of this observation based on the amplitude of the RSB mode will be addressed in Sect. ~\ref{subsec:RSB} ).

Major changes are observed in the vertical plane (Fig.~\ref{fig:VERPIV_AVG}). In particular, there is a clear elongation of the recirculating bubble (see Fig.~\ref{fig:ACx_ACB}c), it can be observed that for arrangements C and V, the recirculation bubble is more asymmetric, whereby the base area facing backflow decreases progressively as $C_q$ increases. These configurations include a vertical plane jet, thus modifying the recirculating bubble more effectively in the vertical direction (a similar effect has been reported for the H configuration in the horizontal plane). Interestingly, for the S configuration, blowing produces a circular dipole for the largest value of $C_q$ analyzed, as shown in Fig.~\ref{fig:VERPIV_AVG}. An asymmetric dipole is also remarked in the horizontal view, presumably distorted by the $N$-state presence. 
Finally, it should be noted that the magnitude of the recirculating velocity decreases when the momentum of the blowing jet is important, e.g. around the threshold $C_q=C^{k}_{q,opt}$, thus altering the balance of fluxes in the recirculating bubble. These changes will be subsequently discussed in terms of relative velocity fields, which will serve as a basis for the analysis of fluxes throughout the recirculating bubble included in Sect.~\ref{sec:fluxes}, following the approach of ~\citet{Khan2024}.   

%-------------------------
% Relative velocity fields
% ------------------------

\begin{figure}[t]
\centering
\includegraphics[width=1\textwidth]{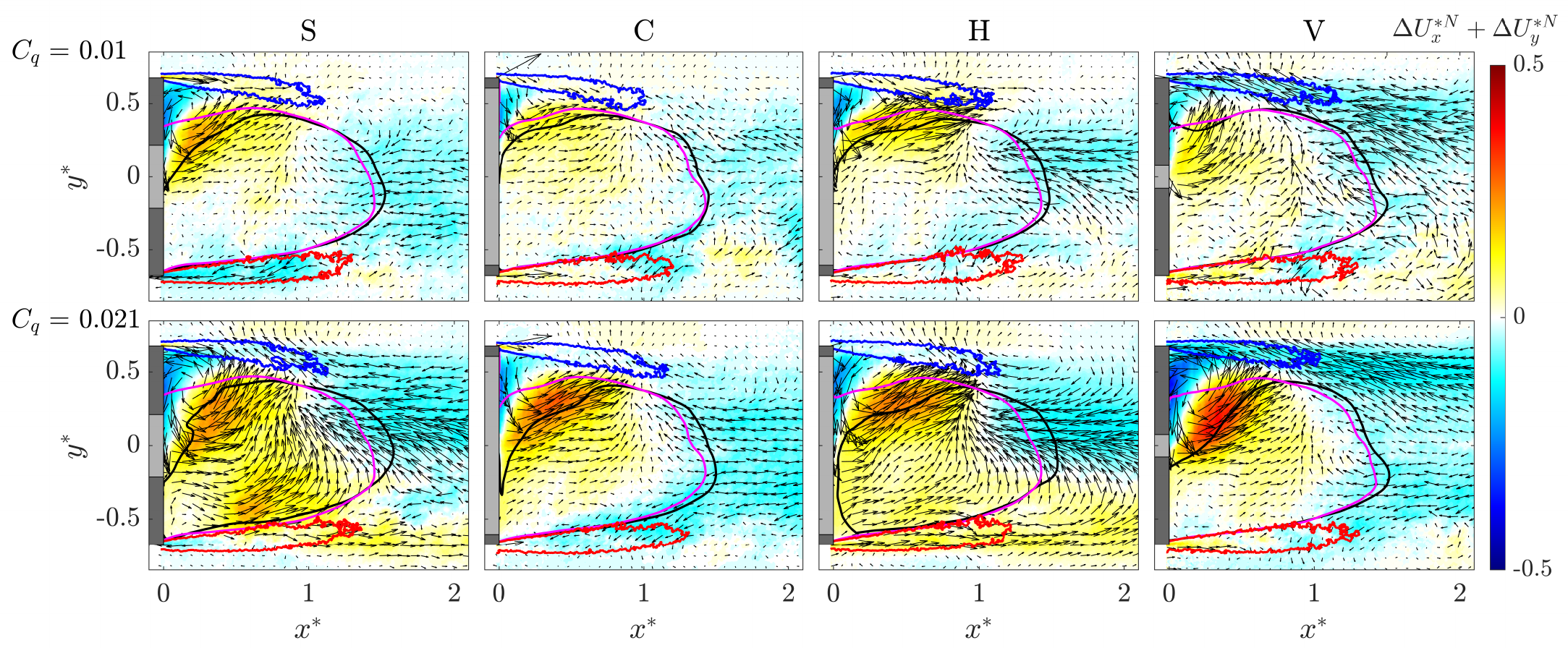}
\caption{\label{fig:DeltaFlow_HOR} Relative flow field between a given blowing case ($k$ = S, C, H, V) and the reference one (B) for the $N$ wake state in the horizontal plane located at $z^*=0$. The contours of the global modifications in both velocity components, $\Delta U^{*N}_{x}+\Delta U^{*N}_{y}$, are also shown. The magenta and black solid lines indicate the boundaries of the recirculation region, defined by the velocity isoline $U^{*N}_x=0$, for the natural case (B) and cases with base blowing, respectively. Finally, the shear layers corresponding to the blowing cases are displayed by $\Omega^{*N}_{z}\pm 4$ vorticity contours for the selected blowing coefficients $C_{q}=[0.01, 0.021]$. The blowing slots are highlighted in light gray in the body base.}
\end{figure}

\begin{figure}[t]
\centering
\includegraphics[width=1\textwidth]{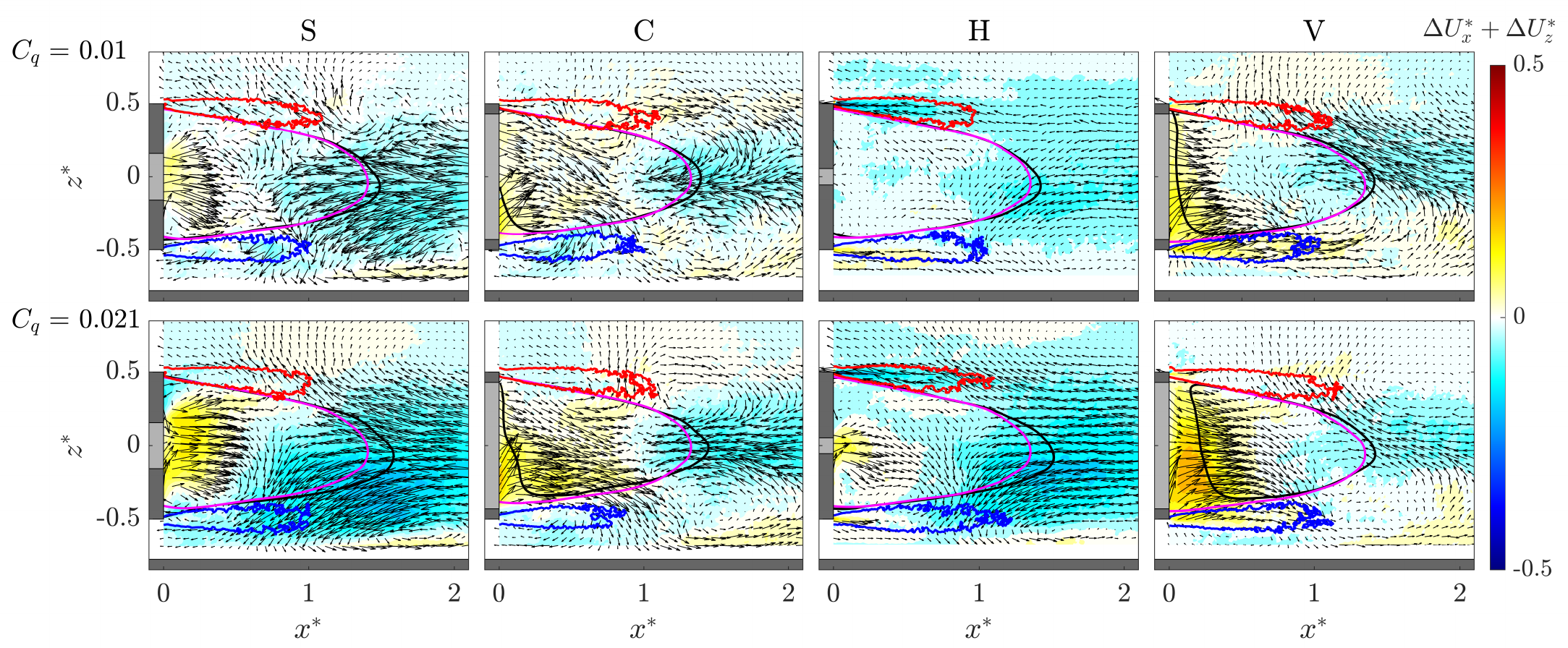}
\caption{\label{fig:DeltaFlow_VER}Relative flow field between a given blowing case ($k$ = S, C, H, V) and the reference one (B) in the vertical plane located at $y^*=0$. The contours of the global modifications in both velocity components, $\Delta U^{*}_{x}+\Delta U^{*}_{z}$, are also shown. The magenta and black solid lines indicate the boundaries of the recirculation region, defined by the velocity isoline $U^{*}_x=0$, for the natural case (B) and cases with base blowing, respectively. Finally, the shear layers corresponding to the blowing cases are displayed by $\Omega^{*}_{y}\pm 4$ vorticity contours for the selected blowing coefficients $C_{q}=[0.01, 0.021]$. The blowing slots are highlighted in gray in the body base.}
\end{figure}

%% DELTA FLOWS
Furthermore, analyzing the relative velocity fields helps to reveal the interaction between the jet and recirculating region. In particular, Figs. \ref{fig:DeltaFlow_HOR} and \ref{fig:DeltaFlow_VER} show respectively the relative velocity fields in the horizontal, and vertical, $y^*=0$, planes for different blowing rates $C_q=(0.01, 0.021)$. These relative flow fields are computed by subtracting the velocity field of the natural case (B) from that obtained when base blowing is applied. The figures illustrate the flow-field modifications introduced by blowing, where $\Delta U^{*}_{x}$, $\Delta U^{*}_{y}$ and $\Delta U^{*}_{z}$ indicate the differences in the $x$, $y$ and $z$ components of the velocity, respectively. They also display contours of the changes produced in both velocity components ($\Delta U^{*N}_{x}+\Delta U^{*N}_{y}$ in the horizontal plane and $\Delta U^{*}_{x}+\Delta U^{*}_{z}$ in the vertical plane) to better illustrate the modifications induced by blowing. A blueish (respectively, reddish) region corresponds to an area where the velocity has decreased (respectively, increased) with respect to the reference area owing to base blowing. A similar procedure to present the effect of blowing devices has been used in \citet{Lorite20} and \citet{Haffner2021}, but using the velocity magnitude instead. This relative flow field is a fair representation of the flow momentum changes induced by the blowing in subsonic flows. Furthermore, the magenta and black solid lines indicate the boundary of the recirculation region for the reference case (B) and cases with blowing, respectively. Finally, the shear layers corresponding to the blowing cases are highlighted by $\Omega^{*}_{i}=\pm 4$ vorticity contours, i.e. $\Omega^{*}_{z}$ or $\Omega^{*}_{y}$ in the selected blowing experiments.

At $C_q=0.01$, in the horizontal plane (shown in Fig~\ref{fig:DeltaFlow_HOR}), the blowing is mainly directed towards the recirculation region located on the $y^{*}>0$ side of the base for the $N$-state. The changes caused by blowing are more obvious for the S and H configurations, while they are less obvious for the C configuration. For the V configuration, with $C_q=0.01$ close to $C^{V}_{q,opt}$, base blowing also modifies the near wake, particularly around the separated shear layers behind the body. In the vertical direction, where the natural flow is almost symmetric, jet injection can easily be observed at $C_q=0.01$ for the S, C, and V configurations (see Fig.~\ref{fig:DeltaFlow_VER}). As discussed earlier, the H configuration does not produce any significant effect on the vertical plane at this blowing flow rate. While the C and V configurations produce planar jets that increase the streamwise velocity throughout the recirculation region, the S arrangement produces a more localized jet. In all cases, the streamwise velocity is reduced at the boundary of the recirculation region owing to an increase in $L^{*}_{r}$.

\begin{figure}[t]
\centering
\includegraphics[width=0.80\textwidth]{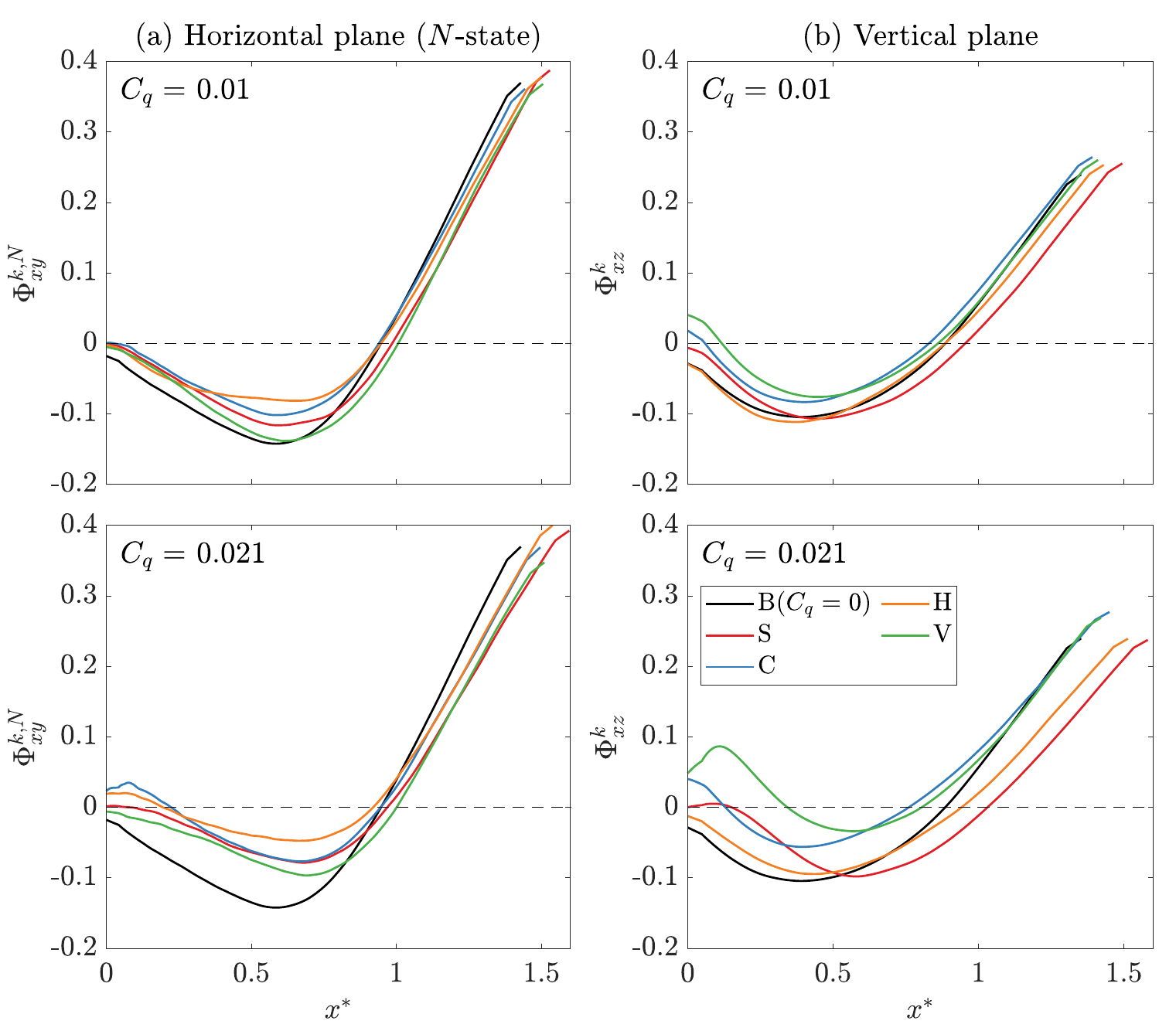}
\caption{\label{fig:Backflow} (a) Recirculation flux, $\Phi^{k,N}_{xy}$, along recirculation region ($0<x^{*}<L^{*}_{r}$), for the $N$-state in the horizontal plane, $z^{*}=0$, for all the configurations at $C_{q}=$ 0.01 and 0.021. The flux was calculated in the range of the body base $-0.5 \, w/h<y^{*}<0.5 \, w/h$ and (b) recirculation flux, $\Phi^{k}_{xz}$, along the recirculation region ($0<x^{*}<L^{*}_{r}$) in the vertical plane, $y^{*}=0$. The flux has been computed in the range of the body base $-0.5 <z^{*}<0.5 $.} 
\end{figure}

However, when the flow rate is further increased to $C_q=0.021$, greater differences between the blowing strategies can be observed. In fact, compared to the low blowing rate case, there is a more significant increase in the recirculation bubble length. At this $C_q$, we are close to the optimal blowing rate $C^{k}_{q,opt}$ for each configuration, and the increase in the recirculation bubble is different for each blowing strategy. The blowing jet is more visible in the recirculation region, and the injected fluid has a higher associated momentum. Because the blowing area in the vertical plane is different, the velocity difference associated with the jet (near the slot location) between the blowing experiments and baseline case is not the same for all configurations. In the S and H configurations, the blown jet effectively fills the entire recirculation bubble in the horizontal plane, reducing the backflow (see Fig. ~\ref{fig:Backflow}a). There is also an effect on the transverse velocity (see the vector directions in Fig.~\ref{fig:DeltaFlow_HOR}), which may explain the reduction in the wake asymmetry. In addition, the S configuration mainly decreases (respectively increases) the flow velocity in the shear layer with negative (respectively positive) vorticity, and symmetrizes the velocity gradients as well. The same behavior, on a smaller scale, occurs under the H configuration. The blown jet in the V configuration, which is already in the momentum regime at $C_q=0.021$, is directed towards the large recirculation zone on the $y^{*}>0$ side of the base for the $N$-state, which increases the asymmetry of the wake. However, the effect of the C configuration is a mixture of the effects observed with the V and H arrangements in the horizontal plane. Figure~\ref{fig:DeltaFlow_VER} shows that in the vertical plane, the flow characteristics at $C_q=0.021$ are similar to those at $C_q=0.01$ but with a greater intensity. Moreover, the S configuration produces an efficient blown jet that reduces the backflow without severely distorting the remaining flow features. Greater flow changes are observed for the C and V configurations. The increase in the recirculation area with the S configuration also causes a reduction in the flow velocity around the downstream stagnation point (as can also be seen in Fig.~\ref{fig:Backflow}b).

%-------------------------
% Relative velocity fields
% ------------------------

\subsubsection{Balance of fluxes in the near wake}\label{sec:fluxes}
%% Recirculating flux
The influence of base blowing on the backflow inside the recirculation region will be quantified by computing the recirculation flux in the horizontal $z^*=0$ plane, for the $N$-state, $\Phi^{k,N}_{xy} (x^*)=\int_{-0.675}^{0.675} U^{*k,N}_{x}(x^{*}) \, dy^{*}$,  and in the vertical $y^*=0$ one, $\Phi^{k}_{xz} (x^*)=\int_{-0.5}^{0.5} U^{*k}_{x}(x^{*}) \, dz^{*}$, for $0\le x^{*} \le L^{*k}_{r}$. These fluxes are calculated for the baseline case (B) and the different blowing configurations ($k$= S, C, H, V) for $C_{q}=$ 0.01 and 0.021, and are represented in Fig.~\ref{fig:Backflow}. In the reference configuration (B), $\Phi^{B,N}_{xy}$ depicts an important backflow, reaching a minimum at approximately $\Phi^{B,N}_{xy}\simeq-0.15$ at $x^{*}\simeq 0.6$ in the horizontal plane. In the vertical plane, the backflow is slightly less intense and the minimum flux is located around $x^{*}\simeq 0.4$. Beyond these points, the fluxes begin to increase as the positive velocity region increases as we move away from the body base.

At $C_{q}=0.01$, the blown jet can produce a slightly positive flux close to the body base for the V and C configurations in the vertical plane, as both configurations produce a vertical planar jet that covers almost the entire body base. At higher blowing flow rates, only C and H configurations are able to produce that effect in the horizontal direction. The downstream evolutions  of $\Phi^{k,N}_{xy}$ and $\Phi^{k}_{xz}$ are similar for the blowing configurations at $C_{q}=0.01$.  The backflow is effectively reduced by V configuration (respectively, H configuration) in the vertical (respectively, horizontal) plane while it is almost unaffected in the horizontal (respectively, vertical) plane. Considering that $\Phi^{k,N}_{xy}$ and $\Phi^{k}_{xz}$ are computed in the interval $0<x^{*}<L^{*}_{r}$, the blowing effect on the recirculation region length can be easily identified in Fig.~\ref{fig:Backflow}. For both the vertical and horizontal planes, the S configuration can further increase the recirculation region at $C_{q}=0.01$, as shown in Figs. \ref{fig:ACx_ACB} - \ref{fig:VERPIV_AVG}. 

At $C_{q}=0.021$, the downstream evolution of the fluxes was strongly affected by blowing in both the horizontal and vertical planes. In the horizontal direction, all the configurations reduce the backflow and shift the minimum flux towards the downstream direction. Since the H configuration consists of a horizontal planar jet, $\Phi^{H,N}_{xy}$ is effectively increased. Similarly, the largest increase in $\Phi^{k}_{xz}$ is observed for the V configuration. The S configuration is able to reduce the backflow and to move the recirculation cores away from the body base, but the greatest change is found in the increase in $L^{*}_{r}$ in both planes, particularly in the vertical one. The slope of the flux after the minimum is also reduced by blowing at $C_{q}=0.021$, reducing the streamwise velocity gradients in the near wake, which is related to the drag reduction. The connection between the reduction in backflow, drag decrease, and attenuation of global unsteady modes was also observed for axisymmetric bodies with a base bleed in \citet{Sevilla04}, \citet{Sevilla06}, and \citet{Bohorquez2011}.

\begin{figure}[t]
\centering
\includegraphics[width=1\textwidth]{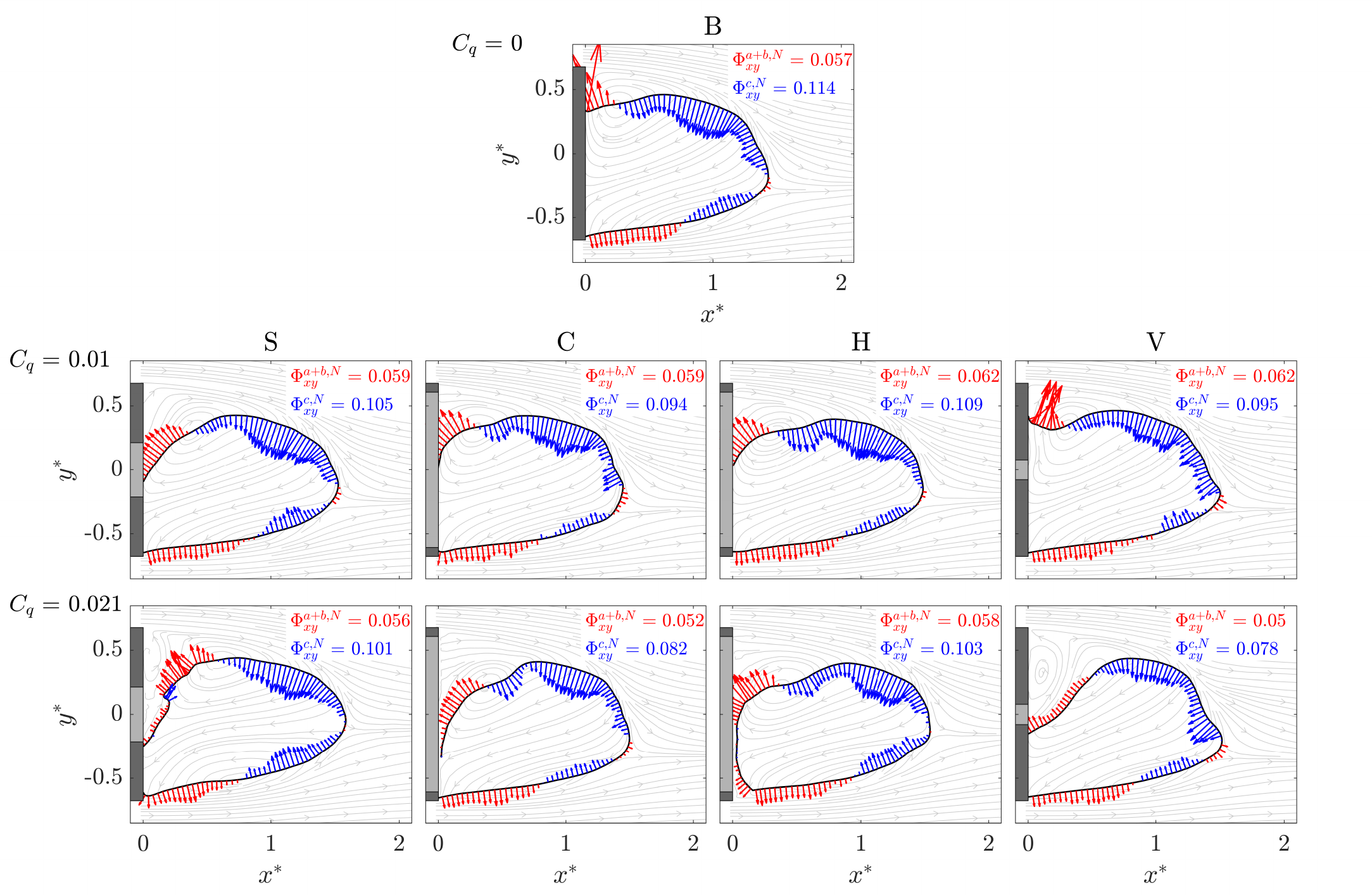}
\caption{\label{fig:HORPIV_RRI}Velocities normal to the boundary of the recirculation region, for the conditionally averaged $N$-state, in the horizontal plane $z^{*}=0$. The reference case (B) and blowing configurations ($k$ = S, C, H, V) are illustrated for $C_{q}=[0.01, 0.021]$. The recirculation region is depicted by the velocity isoline $U^{*N}_{x}=0$. The normal velocity vectors are colored in red for outflows and blue for inflows. The velocity streamlines are depicted in gray in the background in, which illustrates the corresponding near wake topology. The outflow, $\Phi^{a+b,N}_{xy} (k, C_q)$, and inflow, $\Phi^{c,N}_{xy} (k, C_q)$, values are also included in each case.}
\end{figure}

\begin{figure}[t]
\centering
\includegraphics[width=1\textwidth]{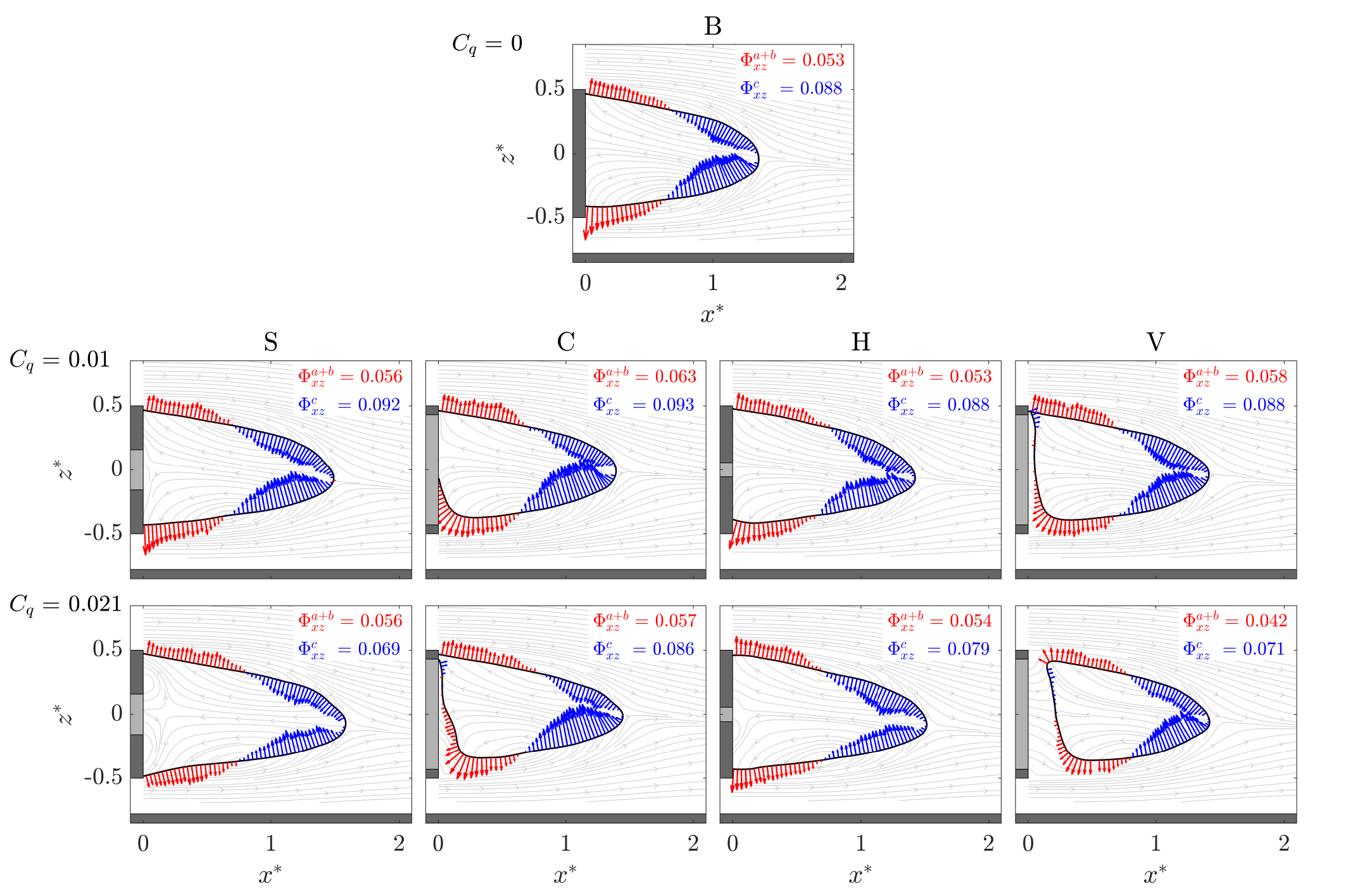}
\caption{\label{fig:VERPIV_RRI}Velocities normal to the boundary of the recirculation region in the vertical plane $y^{*}=0$. The reference case (B) and the blowing configurations ($k$ = S, C, H, V) are illustrated for $C_{q}=[0.01, 0.021]$. The recirculation region is depicted by the velocity isoline $U^{*}_{x}=0$. The normal velocity vectors are colored in red for outflows and blue for inflows. The velocity streamlines are depicted in the gray background, illustrating the corresponding near wake topology. The outflow, $\Phi^{a+b}_{xz} (k, C_q)$, and inflow, $\Phi^{c}_{xz} (k, C_q)$, values are included in each case.}
\end{figure}

%% Fluxes in/out
The equilibrium between the input and output fluxes in the recirculation region has an impact on its length, velocity backflow, and surrounding shear layers \cite{Haffner2021, Khan2024}, affecting the base pressure. Thus, the changes in the recirculating flow are now analyzed based on the fluxes across the recirculation region to understand the role of the blowing geometry on the flow exchanges between the near wake and the unperturbed flow. According to \citet{Gerrard66}, \citet{Stella2018}, \citet{Lorite20}, and \citet{Khan2024}, the key fluxes that control the system are the inflow \(\Phi^c\) and the outflow \(\Phi^a + \Phi^b\). The inflow, or replenishment flow, is responsible for replacing the fluid expelled from the wake region, helping maintain the low-pressure zone near the base of the body. This inflow is typically driven by vortex shedding and occurs downstream of the recirculation region. In contrast, the outflow consists of two main contributions: the vortex roll-up flux \(\Phi^a\), which occurs as vortices form and detach from the recirculation region, and the shear layer entrainment flux \(\Phi^b\), where the fluid is drawn into the wake by the separating shear layers before they roll up into vortices. In addition, base blowing introduces a controlled influx of mass and momentum into the wake through an additional flux \(\Phi^q\). This action alters the natural balance of fluxes by modifying the inflow and outflow rates. Such fluxes are also modified when the length of the recirculation region is altered by blowing devices.

These fluxes are calculated by integrating the velocity normal to the boundary of the recirculation region, defined here by the $U_x^*=0$ isoline, along the boundary.Thus, the aforementioned fluxes are obtained as,

\begin{equation}
\Phi^{a+b}=\int \pmb{U^{*}}\cdot \pmb{n} \, dl^* \rightarrow\pmb{U^{*}}\cdot \pmb{n}\geq0
\end{equation}
\begin{equation}
\Phi^{c}=-\int \pmb{U^{*}}\cdot \pmb{n} \, dl^{*}  \rightarrow\pmb{U^{*}}\cdot \pmb{n}<0
\end{equation}
where $\pmb{n}$ is the normal vector and $dl^{*}$ is the differential length along the boundary of the recirculation bubble.

In the current study, to elucidate the influence of slot geometry on the flux balance of the recirculation region, we propose performing a flux balance analysis similar to that carried out by \citet{Khan2024}. As mentioned in Section \ref{sec:set-up}, the present study is based on two-dimensional PIV measurements performed in two perpendicular planes. Consequently, our measurements do not capture the full three-dimensional nature of the problem under investigation, and the analysis should be understood as a qualitative rather than a quantitative approach.

In this context, flux balance analysis is performed in the horizontal plane ($z^{*}=0$) for the $N$-state of the wake and in the vertical plane at $y^{*}=0$ (see Figs. \ref{fig:HORPIV_RRI}, and \ref{fig:VERPIV_RRI}, respectively). The reference case shows that the fluid enters the recirculation bubble mainly from the front, whereas it exits near the base of the body. In the $N$-state, the main inflow occurs downstream of the large recirculation region located on the $y^*>0$ side of the base. Conversely, the flow balance in the vertical plane is almost symmetrical. The same flow characteristics are observed in the three-dimensional analysis performed by \citet{Khan2024}, confirming our local results. The calculation of the outflows is affected by the resolution of the experimental measurements very close to the base, which reduces the values of $\Phi^{a+b}$ obtained. However, the inflow can be better estimated with our current measurements; therefore, the analysis of the results will mainly focus on this flux.

When blowing is applied, there is an additional inflow in the recirculation region, $\Phi^{q}$. As a result, the replenishment flux, $\Phi^{c}$, must be reduced to compensate for the extra mass added. However, \citet{Khan2024} showed that the mass injected by the blowing jet is much smaller than the $\Phi^{c}$ decrease, suggesting that there is a momentum effect induced by the blowing in the recirculation region. \citet{Khan2024} also reported that this reduction in $\Phi^{c}$ results in elongated shear layers on all sides, which terminate the bubble with a roll-up of reduced intensity at a further downstream location. Therefore, the dominant cause of bubble growth and the accompanying drag reduction is attributed to the momentum of the base blowing.

At $C_q=0.01$, the fluxes are barely modified in comparison with the reference case (see Figs.~\ref{fig:HORPIV_RRI}, \ref{fig:VERPIV_RRI} and Table~\ref{Table_flows}). In the horizontal plane, the largest changes are observed by the modification of the boundary of the recirculation bubble near the large recirculation region on the $y^*>0$ side of the base in the $N$-state. The most significant changes are caused by the stretching of the recirculation bubble during the mass regime. The configurations operating in the horizontal plane (S, C, H) change the curvature of the recirculation bubble boundary near the base on the $y^*>0$ side, thereby modifying the outflow pattern in this region. In the vertical direction, at $C_q=0.01$, only the C and V configurations show a clear modification of the flux exchange compared to the reference case (see Fig.~\ref{fig:VERPIV_RRI}). These variations are introduced by the partial separation of the recirculation area from the base.  

However, the most striking changes in the flux equilibrium are observed at $C_q=0.021$, when the momentum of the jet becomes important and can modify the recirculation region (referred as favorable momentum regime in \citet{Khan2024}). As seen before for the horizontal plane, the S and H configurations can symmetrize the wake and reduce the backflow, resulting in a reduction in $\Phi^{c,N}_{xy}$. The V configuration strongly distorts the near wake even in the horizontal plane, where jet actuation is minimal, as it is already in the momentum regime at $C_q=0.021$. This modifies both $\Phi^{a+b}_{xy}$ and $\Phi^{c}_{xy}$. The flux changes for the C configuration in the horizontal plane are between those for the V and H configurations. In the vertical direction, illustrated in Fig.~\ref{fig:VERPIV_RRI}, the effects of the wake asymmetry are much smaller, with the $\Phi^{a+b}_{xz}$ and $\Phi^{c}_{xz}$ fluxes better estimated in this plane. In any case, we must remember that we are carrying out a two-dimensional analysis, which may be lacking in some relevant three-dimensional information. The S and V configurations are capable of significantly reducing $\Phi^{c}_{xz}$, decreasing the induced backflow. This strong reduction of the replenishment fluid flux under S arrangement, might be linked with the maximum increase in the recirculation length $L_r^*$ reported in Fig. \ref{fig:ACx_ACB}(c) and Fig. \ref{fig:Backflow}(b). In addition, under S configuration, the magnitude of the normal velocity over the recirculation region is reduced for both the inflow and outflow, leadin to a decrease in the velocity gradients in the near wake. As a result, the amplitude of the flow fluctuations associated with vortex shedding are reduced, which explains the greatest drag reduction observed, similar to the results reported in \cite{Khan2024}. Since the V configuration introduces a plane jet in the vertical direction, which detaches the recirculation bubble from the base, the outflow $\Phi^{a+b}_{xz}$ also decreases. A similar feature is observed for the H configuration in the horizontal direction. However, the H configuration globally increases more efficiently the recirculation length, which in turn reduces the normal velocity across the recirculation region, resulting in an almost constant $\Phi^{a+b}_{xy}$. Once again, the C configuration shows an intermediate scenario between V and H, as it was the case in the previous results.
\begin{table}[t]
\label{Table_flows}
\centering
\begin{tabular}{@{}cc|cccc|cc@{}}
\toprule
$C_q$ &
  Conf &
  $\Delta \Phi^{a+b,N}_{xy}(\%)$ &
  $\Delta \Phi^{c,N}_{xy}(\%)$ &
  $\Delta \Phi^{a+b}_{xz}(\%)$ &
  $\Delta \Phi^{c}_{xz}(\%)$ &
  $\langle TKE^{N}_{xy} \rangle$ &
  $\langle TKE_{xz} \rangle$ \\ \midrule
                         & S & 2.18                         &-7.59                          & 6.74  & 3.86   & \cellcolor[HTML]{FFFFFF}0.169 & 0.207 \\
                         & C & 3.27                          &-17.06                         & 19.25 & 5.39   & \cellcolor[HTML]{FFFFFF}0.172 & 0.207 \\
                         & H & 7.58                          & -4.39                         & 0.82  & -0.10  & \cellcolor[HTML]{FFFFFF}0.170 & 0.214 \\
\multirow{-4}{*}{0.010} & V & 7.99                         & -16.27                        &  9.49 & -1.51  & \cellcolor[HTML]{FFFFFF}0.173 & 0.206 \\ \midrule
                         & S & \cellcolor[HTML]{FFFFFF}-2.01  & \cellcolor[HTML]{FFFFFF}-11.52  & 6.35 & -21.43 & \cellcolor[HTML]{FFFFFF}0.166 & 0.204 \\
                         & C & \cellcolor[HTML]{FFFFFF}-10.17 & \cellcolor[HTML]{FFFFFF}-27.74 & 8.31 & -2.62  & \cellcolor[HTML]{FFFFFF}0.163 & 0.200 \\
                         & H & \cellcolor[HTML]{FFFFFF} 0.13  & \cellcolor[HTML]{FFFFFF}-9.81 & 1.88  & -10.20 & \cellcolor[HTML]{FFFFFF}0.162 & 0.211 \\
\multirow{-4}{*}{0.021} &
  V &
  \cellcolor[HTML]{FFFFFF}-13.85 &
  \cellcolor[HTML]{FFFFFF}-30.97&
  -20.65 &
  -20.21 &
  \cellcolor[HTML]{FFFFFF}0.174 &
  0.194 \\ \bottomrule
\end{tabular}
\caption{Relative changes of the outflow, $\Delta \Phi^{a+b}_{ij}(k, C_q)$, and inflow $\Delta \Phi^{c}_{ij} (k, C_q)$, across the boundaries of the recirculating bubble for the different configurations ($k$ = S, C, H, V) at $C_{q}$= [0.01, 0.021] in the horizontal ($ij=xy$) and vertical ($ij=xz$) planes. The last two columns show the spatially averaged turbulent kinetic energy $\langle TKE_{ij}\rangle$ in the measurement regions depicted in Figs. \ref{fig:HORPIV_k}, \ref{fig:VERPIV_k} for both the horizontal ($ij=xy$) and vertical ($ij=xz$) planes. The reference values for the fluxes are $\Phi^{a+b,N}_{xy}$(B)$=0.057\pm 0.005$, $\Phi^{c,N}_{xy}$(B)$=0.114\pm 0.005$, $\Phi^{a+b}_{xz}$(B)$=0.053\pm 0.002$, $\Phi^{c}_{xz}$(B)$=$ $0.088$ $\pm 0.002$, and for the turbulent kinetic energy they are $TKE^{N}_{xy}$ (B) $=0.180$ $\pm 0.006$, $TKE_{xz}$(B)$=0.219\pm 0.001$.}
\end{table}

%% TABLE FLUXES
To further analyze the effect of blowing on the balance of fluxes, we compare the values of $\Phi^{a+b}_{ij}$ and $\Phi^{c}_{ij}$ with the baseline case in Table \ref{Table_flows}, providing $\Delta \Phi^{a+b}_{ij} (\%) = (\Phi^{a+b}_{ij}(k, C_q)-\Phi^{a+b}_{ij}$(B)$)/\Phi^{a+b}_{ij}$(B) and $\Delta \Phi^{c}_{ij} = (\Phi^{c}_{ij}(k,C_q)-\Phi^{c}_{ij}(B))/\Phi^{c}_{ij}$(B). This relative variation was calculated in two measured planes: the horizontal plane for the $N$-wake state denoted by the superscript $ij=xy$ and the vertical plane denoted by the superscript $ij=xz$. A positive (negative) variation indicates an increase (decrease) in the flux for the specified $C_q$ with respect to the case without blowing $C_q=0$. Moreover, the dataset incorporates the spatially averaged turbulent kinetic energy values within the near wake region, which are discussed in Section~\ref{sec:tke}.

For $C_q = 0.01$, all the blowing configurations slightly increase the outflow in the horizontal plane $\Phi^{a+b,N}_{xy}$ while moderately decreasing the inflow $\Phi^{c,N}_{xy}$. This is consistent with the mass regime in which the momentum fluxes in the recirculation region remain almost unperturbed. The V configuration produces the largest modifications at $C_q = 0.01$, because it reaches $C_{q,opt}$ at lower blowing rates. It also modifies the fluxes in the horizontal plane, even though the actuation is small in that plane. Conversely, the H configuration does not change the fluxes in the vertical plane at $C_q = 0.01$. For $C_q = 0.021$, when the momentum of the base blowing is larger, the changes in the flux exchange over the recirculation region are more important. The variations of the outflow and the inflow take place in the horizontal as well as in the vertical plane, almost independent of the blowing geometry. As mentioned earlier, the V configuration produces the largest changes because it is already in the momentum regime. The S arrangement is  the most efficient configuration for reducing the inflow in the vertical direction, $\Phi^{c}_{xz}($S$, C_{q}=0.021)$, whereas modifications in the horizontal direction are affected by the symmetrization of the wake.

%% TKE 
\subsubsection{Effect of blowing in the wake fluctuations}\label{sec:tke}
\begin{figure}[t]
\centering
\includegraphics[width=1\textwidth]{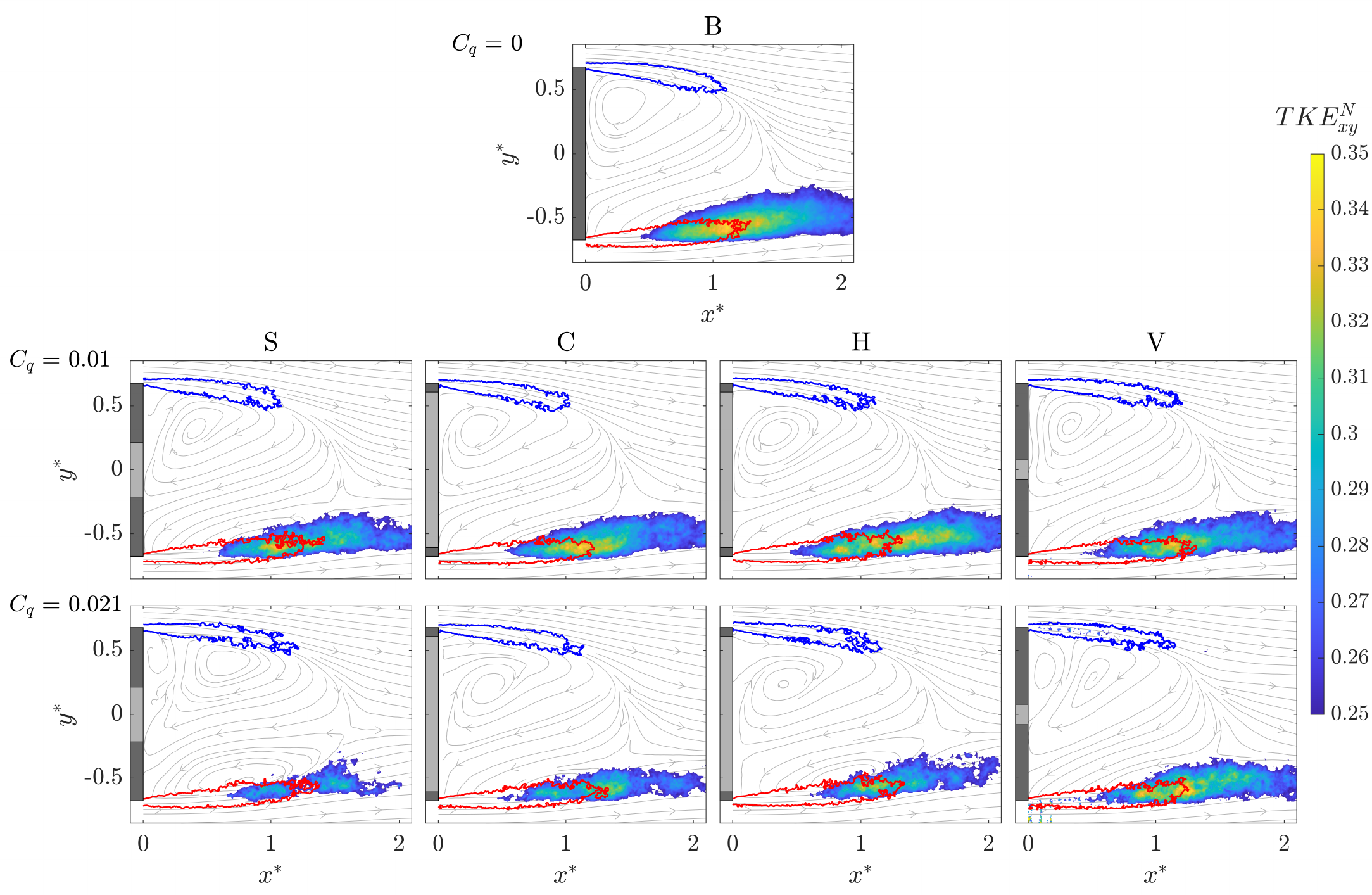}
\caption{\label{fig:HORPIV_k} Conditionally averaged turbulent kinetic energy contours, $TKE^N_{xy} > 0.25$, for $N$-state and vorticity isocontours, $\Omega^{*N}_{z}=\pm 4$, for the reference (B) and the different blowing configurations ($k$ = S, C, H, V) with $C_{q}=[0.01, 0.021]$ in the horizontal $z^{*}=0$ plane.}
\end{figure}
\begin{figure}[t]
\centering
\includegraphics[width=1\textwidth]{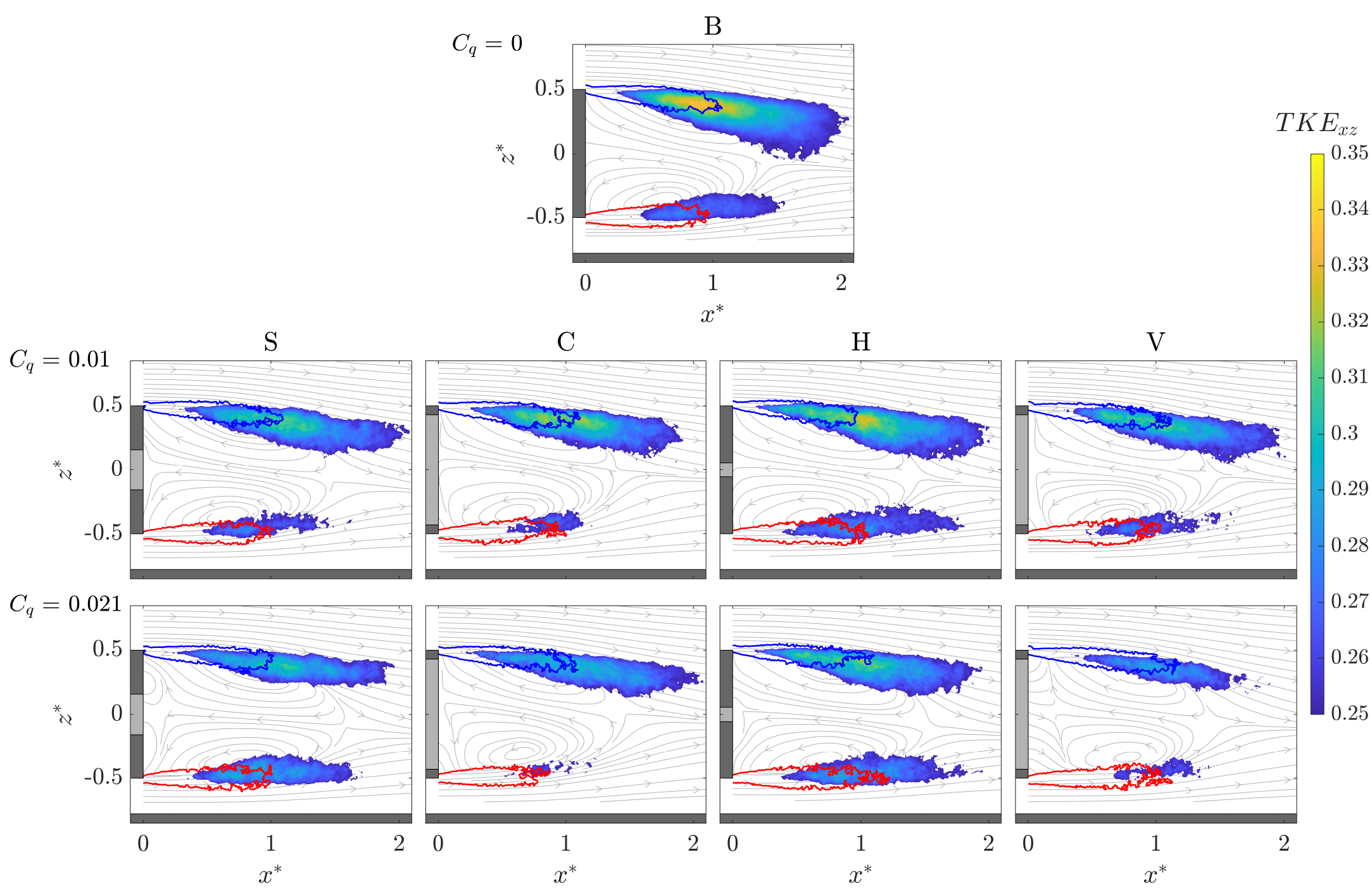}
\caption{\label{fig:VERPIV_k} Averaged turbulent kinetic energy contours, $TKE_{xz} > 0.25$, and iso-surfaces of vorticity, $\Omega^{*}_{y}=\pm 4$, for the reference case (B) and the different blowing configurations ($k$ = S, C, H, V) with $C_{q}=[0.01, 0.021]$ in the vertical $y^{*}=0$ plane.}
\end{figure}

The changes observed in the near wake with base blowing, i.e. the increase in $L_r^*$, decrease in backflow, and overall decrease in $\Phi^c$, are related. These changes modify the velocity distributions in the near wake and turbulence production in the shear layers surrounding the recirculation region. As reported in ~\citet{Haffner2020} and \citet{Khan2024}, a more elongated recirculation region in combination with a less intense backflow, due to a smaller inflow, leads to a reduction in the turbulent kinetic energy production, velocity gradients, and velocity fluctuations in the shear layers. In this regard, Figs. \ref{fig:HORPIV_k} and \ref{fig:VERPIV_k} display the turbulent kinetic energy distribution in the horizontal and vertical planes, respectively, for the reference case (B) and the different blowing configurations for $C_{q}=[0.01, 0.021]$. The velocity fluctuations are only shown for $TKE$ values above 0.25 to better illustrate the effect of blowing. The shear layers are also depicted by the vorticity isocontours, already employed in Fig.~\ref{fig:PIV_Gy_Gz_Ref}, $\Omega^{*}_{z}=\pm 4$ and $\Omega^{*}_{y}=\pm 4$. As shown in Fig.~\ref{fig:HORPIV_k}, the S and H configurations are the most efficient in terms of reducing flow unsteadiness in the horizontal plane and symmetrizing the $N$-state of the wake, especially for $C_q=0.021$.
The flow unsteadiness distribution in the vertical plane for the reference case shows a slight asymmetry due to the presence of the ground. The blowing reduces globally the level of fluctuations, especially for $C_q = 0.021$ (see Fig.~\ref{fig:VERPIV_k}). The slight vertical asymmetry of the flow fluctuations is effectively reduced by the S configuration for $C_q = 0.021$. As previously mentioned, the H configuration does not produce any significant change in the vertical direction for the studied blowing flow rates. However, the planar jets created by the V and C setups can interact with the shear layers and begin to induce some asymmetry in the recirculation region, as shown in Fig.~\ref{fig:VERPIV_AVG}. This imbalance causes the shear layer  with positive vorticity to shorten (red) and that with negative vorticity to lengthen (blue). The spatially averaged values of $TKE$ in the entire measurement domain are depicted in Figs. \ref{fig:HORPIV_k} and \ref{fig:VERPIV_k} are also included in Table \ref{Table_flows} to quantitatively measure the reduction in the turbulence production with blowing. As can be seen, blowing within the mass regime does not significantly modify the $TKE$ values in any of the planes for the tested configurations. The largest differences are found for $C_q=0.021$, close to the optimal blowing, for which the S and H configurations are able to efficiently reduce the velocity fluctuations in the horizontal plane (see Fig.~\ref{fig:HORPIV_k}) and the V configuration in the vertical one (see Fig.~\ref{fig:VERPIV_k}). The S configuration increases the level of fluctuations in the vertical direction, which can increase drag; however, this configuration can vertically symmetrize the wake, unlike the C and V arrangements (see Fig.~\ref{fig:VERPIV_k}). As a result, the jet suppresses wake fluctuations in the plane where it is injected. Accordingly, vertically distributed slots (V and C) reduce the turbulent kinetic energy ($TKE$) in the vertical plane, whereas horizontally distributed slots (S and H) have a similar effect in the horizontal plane. Given that the body under study is wider than it is tall ($w>h$), it can be assumed that the global turbulence is more effectively reduced by the slots acting in the horizontal plane (which has a larger surface area than the vertical plane), namely, the S and H configurations. The reduction of the turbulent production is known to contribute to reduce the drag in road vehicles \citep{Bonnavion2022} and confirms the drag trends shown in Fig.~\ref{fig:ACx_ACB}.
  
\subsubsection{Impact of the blowing on the RSB mode}\label{subsec:RSB}

\begin{figure}[t]
\centering
\includegraphics[width=1\textwidth]{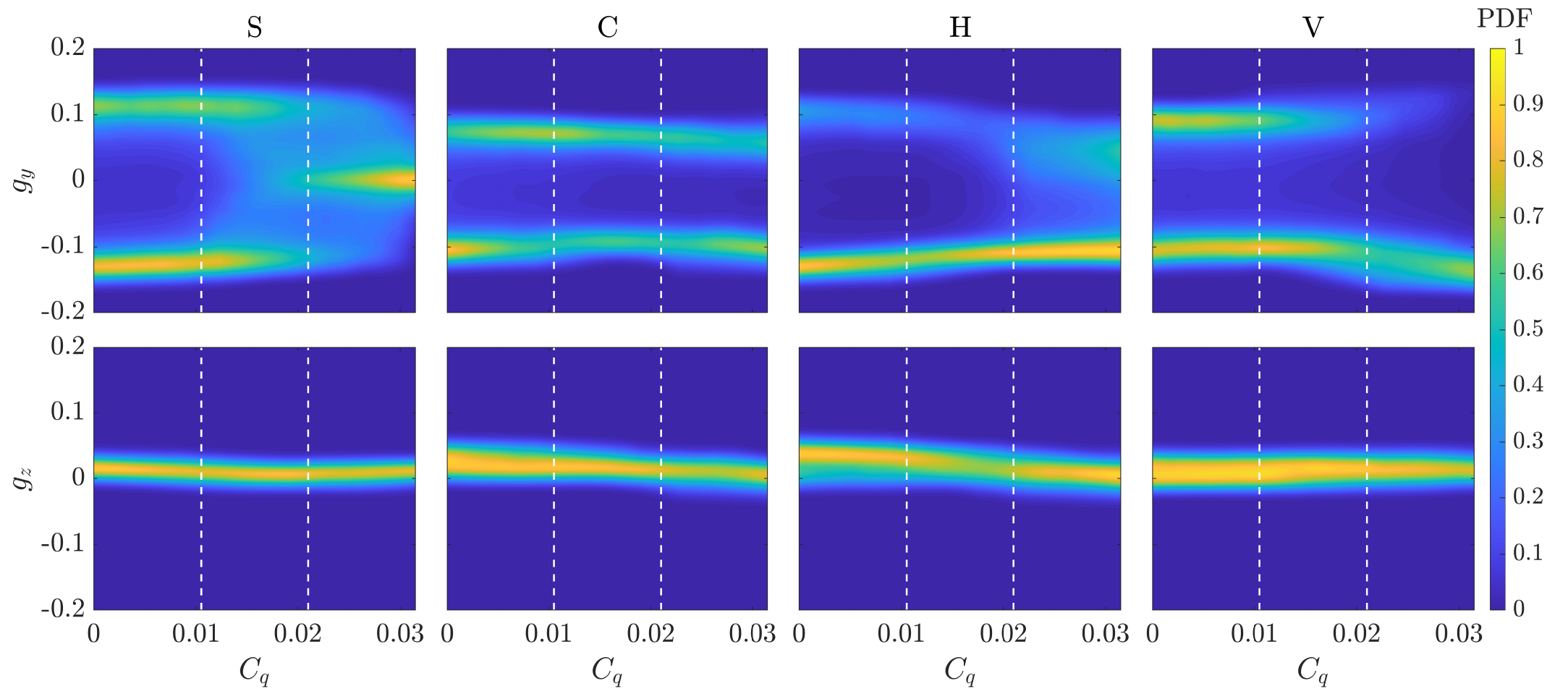}
\caption{\label{fig:Contour_GyGz} Probability Density Functions (PDFs) of the horizontal, $g_y$, and vertical, $g_z$, base pressure gradients for the different blowing configurations (S, C, V, H) as a function of the flow rate, $C_{q}$. The dashed white lines represent the selected values of $C_q$ for which the PIV measurements were performed in both the vertical and horizontal planes. All PDFs are normalized with their corresponding maxima.}
\end{figure}

In the previous sections, we have qualitatively discussed the effect of blowing on the near wake asymmetry by describing the modified recirculation region flow characteristics and the effect of base blowing on them (see Figs. \ref{fig:HORPIV_AVG}-\ref{fig:VERPIV_k}). In this section, we quantitatively analyze the effect of blowing on the near wake asymmetry by focusing on the evolution of the horizontal and vertical base pressure gradients with increasing flow rates. Thus, Fig.~\ref{fig:Contour_GyGz} illustrates the Probability Density Functions (PDFs) of the horizontal base pressure gradient $g_y$ (top row) and the vertical base pressure gradient $g_z$ (bottom row) for the different blowing configurations over the tested flow rates $C_q$. 

The contours of the PDF($g_y$) reveal the initial bi-stability of the RSB mode without blowing, indicating two equally probable wake states: quasi-steady asymmetric states with positive ($P$-state) or negative ($N$-state) values of $g_y$, as shown in Fig.~\ref{fig:PIV_Gy_Gz_Ref}. The initial wake deflection states for all configurations have the same asymmetry as the reference case. However, the V and C arrangements show smaller amplitudes of the horizontal base pressure gradient for their respective $P$ and $N$ states, due to the different distributions of pressure taps used in our experiments (see Fig.~\ref{fig:setup}). 
Note that these small differences in the initial asymmetry values are not significant for our analysis, as we focus on how the blowing configurations globally affect wake symmetrization. 

The PDFs of $g_y$ indicate that the wake exhibits bi-stability in the horizontal direction when $C_q<C_{q,opt}^k$, confirming that, in the mass regime, the blowing acts as a passive scalar in the recirculation region, implying that it lacks sufficient momentum to affect the RSB mode. Only a slight decrease in the asymmetric amplitude was noted by \citet{Lorite20}. In the vertical direction, the mass regime blowing rates $C_q<C_{q,opt}^k$ can only minimally mitigate the initial vertical asymmetry in the pressure gradient $g_z$ caused by the presence of the ground, as shown in the wake in Figs.~\ref{fig:HORPIV_AVG}-\ref{fig:VERPIV_k}. At blowing flow rates close to $C_{q,opt}^k$, the sufficient momentum from the base blowing alters the near wake asymmetry differently. The S and H slots, which primarily affect the horizontal direction, significantly reduce horizontal wake asymmetry. In particular, the S configuration symmetrizes the near wake, as indicated by the PIV measurements (see Figs. ~\ref{fig:HORPIV_AVG}–\ref{fig:VERPIV_k}), and reaches base pressure gradient values close to zero (see Fig.~\ref{fig:Contour_GyGz}). Suppression of the RSB mode with a central blowing slit has also been reported by \citet{Khan2022, Khan2024}, although in their study they found a minimum asymmetry at $C_q<C_{q,opt}$, in contrast to our results.  The use of a smaller blowing slot in their work results in a stronger blown jet momentum, allowing for symmetrization of the RSB mode at lower $C_q$ values, as discussed in detail in  Sect. \ref{sec:Discussion}. The significant reduction in wake asymmetry for configurations S and H during the momentum regime is likely to explain the small slopes of $\Delta C_x$ and $\Delta C_B$ versus $C_q$ for these configurations (see Fig.~\ref{fig:ACx_ACB}) because wake asymmetry is a recognized source of drag \cite{Bonnavion2018, Haffner2021}.  In the C arrangement, bi-stability is maintained over all tested values of \(C_q\). The symmetric distribution of the blown jet does not affect the symmetries of the natural flow; therefore, the initial asymmetry near the wake is maintained. A similar finding was reported by \citet{Lorite20} for symmetric perimetric configurations called TB (Top-Bottom slits) and LR (Left-Right slits). In contrast, the V configuration mainly affects the vertical direction, resulting in greater horizontal wake asymmetry for a constant $g_z$. In this arrangement, increased blowing amplifies the horizontal asymmetry when $C_q > C_{q,opt}$, preserving the bistable behavior of the RSB mode, and eventually establishing an $N$ state with increased asymmetry, as shown in Figs.~\ref{fig:HORPIV_AVG}-\ref{fig:VERPIV_k}. The increased asymmetry may affect the stepper slopes of $\Delta C_x$ and $\Delta C_B$ in the momentum regime for the V slot. The results indicate a strong connection between near wake asymmetry and evolution of $C_x$ and $C_B$ in the momentum regime, which will be discussed in Sect. \ref{sec:Discussion}.

 %%%%%% CONCLUSIONS %%%%%%
\section{Discussion}\label{sec:Discussion}
The observed hierarchy of drag reduction between the tested configurations is based on the interaction of three key factors: bubble elongation, flux modification across the recirculating bubble associated with reduced turbulence generation, and symmetrization of the RSB mode. In this section, we compare our results with those of previous similar studies to shed light on the effects of the different blowing configurations tested here on these factors.

\begin{figure}[t]
\centering
\includegraphics[width=0.75\textwidth]{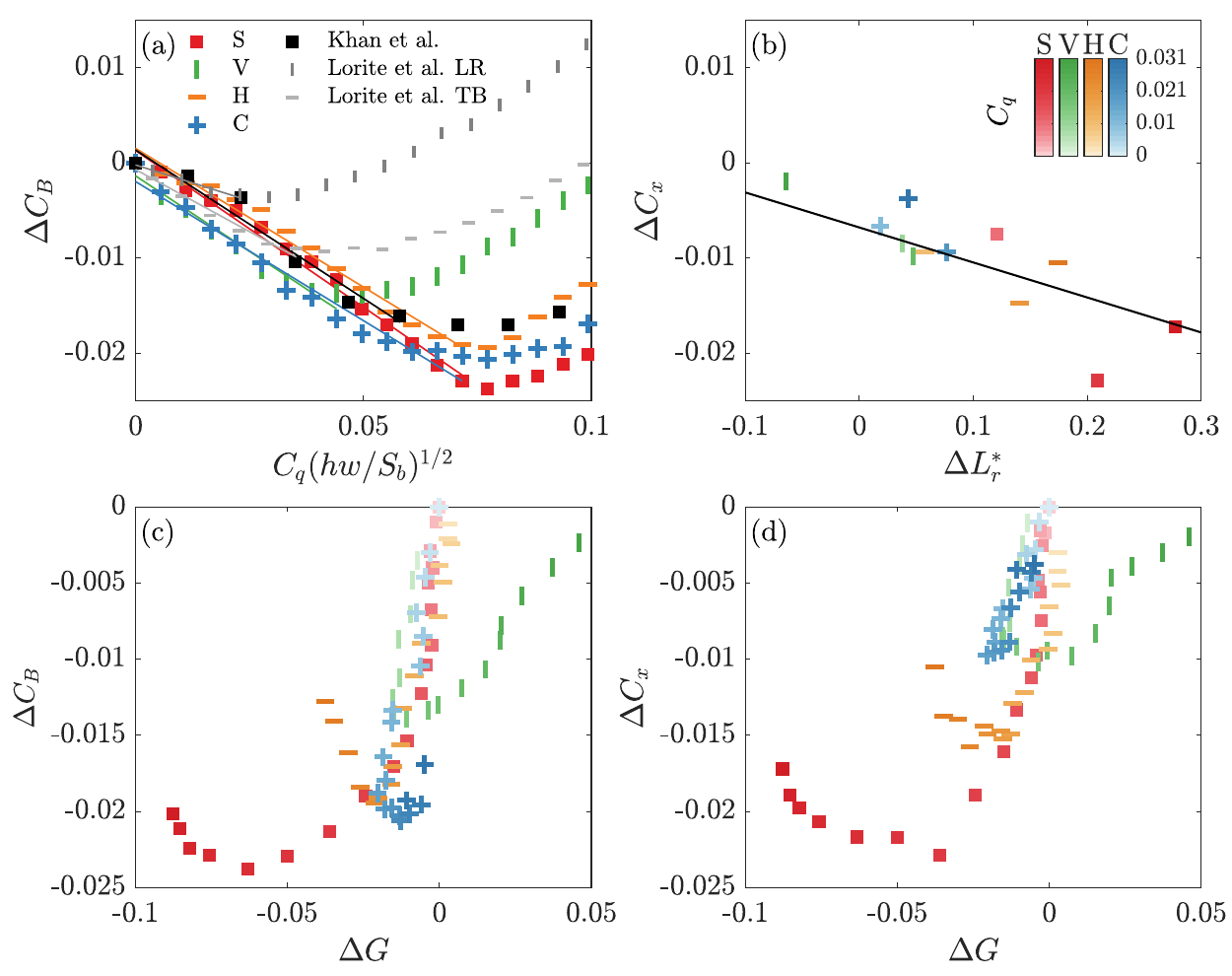}
\caption{\label{fig:Discussion} (a) Base drag reduction, $\Delta C_B$, versus the area-scaled flow rate, $C_q(hw/S_b)^{1/2}$, where the solid lines indicate the slope in the mass regime. (b) Evolution of changes in the dag, $\Delta C_x$, with changes in the length of the recirculation bubble, $\Delta L_r^*$. (c) Correlation between base drag coefficient variation $\Delta C_B$ and total base pressure asymmetry variation  $\Delta G$, and (d) correlation between drag coefficient variation $\Delta C_x$ and total base pressure asymmetry variation  $\Delta G$. %= G^k(C_q)- G^B$, where $G^B = 0.122 \pm 0.001$. 
Note that symbols in (b-d) are colored by the values of $C_q$ following the colormap illustrated in (b).}
\end{figure}

Our study identifies the mass and momentum regimes described by \citet{Lorite20}, including their combined behavior when $C_q \approx C_{q,opt}$, which is consistent with the recently labeled \textit{favorable momentum} regime by \citet{Khan2024}. However, a clear transition between the mass and favorable momentum regimes cannot be clearly identified in the present analysis. For $C_q<C_{q,opt}$, we observe an approximately linear reduction in $C_x$ and $C_B$ with $C_q$, as shown in Fig.~\ref{fig:ACx_ACB}(a,b). However, \citet{Khan2022} and \citet{Khan2024} identified a dual trend in base pressure recovery, indicating a shift between the mass and favorable momentum regimes. This discrepancy between similar studies is attributed to the momentum of the blown jets, which depends on the different sizes of the blowing slits used in each study. These variations affect the interaction between the near wake and base blowing. For this reason, \citet{Khan2022} proposed a scaling method based on a momentum limit for drag reduction induced by central blowing slots of different sizes $C_q(hw/S_b)^{1/2}$. The scaling was applied, as shown in Fig.~\ref{fig:Discussion}(a) to compare our results with those of \citet{Lorite20} and \citet{Khan2022}, who studied symmetric blowing devices acting on the wake of a square base Ahmed body under similar flow conditions. In both cases, the wake exhibited the RSB mode in the horizontal plane for the reference case without blowing. All cases reveal a common trend $\Delta C_{x} = \Delta C_{B} \simeq -0.3 \, C_q(hw/S_b)^{1/2}$ during the mass regime, as observed in Fig.~\ref{fig:Discussion}(a) and Table \ref{tab:Discussion}. In addition, for blowing devices with a central slot away from the body edges, the optimal value $C_{q,opt}(hw/S_b)^{1/2}$ of the minimum drag is the same because the transition mechanism between the mass and momentum regimes remains unchanged.  

When the central jet momentum is sufficiently large to severely disrupt the backflow (or the flow towards the body base), drag reduction ceases. The hypothesis of a limit in the blown momentum is supported by common scaling with $C_{q}(hw/S_b)^{1/2}$. Applying the same scaling to perimetric blowing data, where an increased $\Phi^b$ facilitates blowing escape from the recirculation region through shear layers, a limited drag reduction is obtained, and $C_{q,opt}(hw/S_b)^{1/2}$ decreases. In our study, intermediate configurations between perimetric and central blowing slits are tested. The H, V, and C arrangements also follow evolution $\Delta C_{x} = \Delta C_{B} \simeq -0.3 \, C_q(hw/S_b)^{1/2}$. However, the planar vertical jet from V configuration escapes the recirculation region easily which decreases the efficiency of the blowing in that arrangement. % because the width of the body exceeds its height, as $w=1.35\, h$ (note that $C^{H}_{q,opt}(hw/S_b)^{1/2}\simeq 1.35 \, C^{V}_{q,opt}(hw/S_b)^{1/2}$~). 
The good performance of the H configuration is also linked to the presence of the RSB mode in the horizontal direction of the natural wake, which minimizes the modification of the near wake with horizontal blowing. Based on this analysis, we conjecture that the evolution of the drag with base blowing is related to the following factors: 

\begin{itemize}
    \item \textbf{Recirculation bubble elongation}: The stable, low-momentum jet fills the recirculation bubble, increasing its length and moving the recirculation core away from the body, restoring base pressure and reducing the drag. In this regard, Fig.~\ref{fig:Discussion}(b) illustrates the relationship between $\Delta C_{x}$ and the change in the length of the recirculation region $\Delta L^{*}_r$. For the tested geometries, the drag reduction exhibits a consistent trend during the mass regime (see solid line in fig.~\ref{fig:Discussion}b), suggesting that slot geometry does not play a major role in the drag reduction mechanism during the mass regime. However, it significantly influences the critical flow rate $C^{k}_{q,opt}$ and the characteristics of the momentum regimes. 
    
    \item \textbf{Fluxes balance and wake turbulence}: Base blowing around $C^{k}_{q,opt}$ alters the inflows and outflows through the recirculation region, thereby modifying the backflow within the recirculation region and the velocity gradients in the massively separated flow (see Figs. \ref{fig:HORPIV_AVG},-\ref{fig:VERPIV_RRI}). These modifications affect the turbulence intensity caused by the velocity fluctuations in the near wake and drag (see Figs. \ref{fig:ACx_ACB}, \ref{fig:HORPIV_k}, \ref{fig:VERPIV_k}). Recently, \citet{Khan2024} identified an additional drag reduction using this mechanism for a central slit. Our  Fig.~\ref{fig:Discussion}(b) shows a similar deviation in the relationship between $\Delta C_{x}$ and $\Delta L^{*}_r$ for the S configuration near the optimum blowing rate. This increased drag reduction is driven by the base blowing momentum, and therefore, the slit geometry primarily determines the critical blowing flow rate, $C^{k}_{q,opt}$, and the maximum achievable drag reduction. In fact, a larger blowing slot, such as the one used here, combined with lower blowing velocities, reduces the drag more effectively than that observed in previous studies (see Fig.~\ref{fig:Discussion} a, Table \ref{tab:Discussion}). This is probably because the lower jet momentum promotes passive scalar behavior, allowing efficient recirculation bubble filling over a wider range of blowing coefficients.
     
    \item {\textbf{Wake asymmetry}: The wake asymmetry is primarily driven by the RSB mode present in the horizontal plane and, to a lesser extent, by the pressure gradient in the vertical plane due to the presence of the ground (see Table~\ref{tab:RefData}). Next, we discuss the effect of blowing on global asymmetry using Fig. ~\ref{fig:Discussion} (c, d), which plots the changes in the total average base pressure asymmetry $\Delta G= G^k(C_q)- G^B$ (where $G^B = 0.122 \pm 0.001$) versus $\Delta C_B$ (Fig.\ref{fig:Discussion}c) and $\Delta C_x$ (Fig.\ref{fig:Discussion}d) respectively. On the one hand, the effect of increasing blowing (illustrated by changes in color intensity) during the mass regime has a limited effect on the wake, reducing $G$ by only $10\%-20\%$, as in \citet{Lorite20}, while the bistable behavior of the wake persists for all blowing configurations, as discussed in Sect.~\ref{subsec:RSB}. Nevertheless, increasing the blowing within the momentum regime significantly alters the wake asymmetry and can suppress the RSB mode depending on the slit configuration. The central slit configurations from different studies exhibit nearly identical $C_{q,opt}(hw/S_b)^{1/2}$ values for the optimum drag reduction (Fig.~\ref{fig:Discussion} a). However, the RSB mode was suppressed at $C_{q}(hw/S_b)^{1/2} \simeq 0.021$ for small slots with $S_{b}/hw \simeq 0.0057$ in \citet{Khan2022, Khan2024} while in our work ($S_{b}/hw = 0.1$), the RSB suppression occurs at $C_{q}(hw/S_b)^{1/2} \simeq 0.1$. This suggests that factors other than jet momentum are likely to contribute to the attenuation of the RSB mode; however, further investigation is beyond the scope of this manuscript. For $C_q > C^{k}_{q,opt}$, the drag reduction ceases, and the wake asymmetry is significantly affected, depending on the characteristics of the blowing slot.  \citet{Haffner2020} showed that increased wake asymmetry increases the shear layer interaction, which increases drag. Consequently, drag increases slightly with blowing for configurations such as S and H, which minimize wake asymmetry in the momentum regime. This occurs because the negative effects of the other two aforementioned factors are partly offset by the drag reduction achieved through wake symmetrization. Conversely, if the wake asymmetry is promoted during the momentum regime, as in the V setup, the drag is largely increased for $C_q>C_{q,opt}$.}  
\end{itemize}

\begin{table}[t]
\centering
\begin{tabular}{ccc|cc}
\hline
\multicolumn{3}{c|}{Slot}   & \multicolumn{2}{c}{Results} \\ \hline
Study     & Geometry & $S_{b}/hw$   & $\partial \Delta C_B/\partial(C_q(hw/S_b)^{1/2})$   & $\Delta C_{B,max}$  \\ \hline
Present (S)&      \begin{minipage}{.028\textwidth}
      \includegraphics[width=0.9\linewidth]{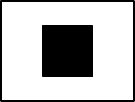}
    \end{minipage}        & 0.1    & -0.323   & -0.0238  \\
Present (V) &  \begin{minipage}{.028\textwidth}
      \includegraphics[width=0.9\linewidth]{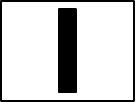}
    \end{minipage}        & 0.1    & -0.317   & -0.0139  \\
Present (H) &   \begin{minipage}{.028\textwidth}
      \includegraphics[width=0.9\linewidth]{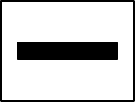}
    \end{minipage}       & 0.1    & -0.292   & -0.0194  \\
Present (C) &   \begin{minipage}{.028\textwidth}
      \includegraphics[width=0.9\linewidth]{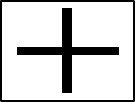}
    \end{minipage}       & 0.1    & -0.293    & -0.0206  \\
Khan et al. [2022]    &   \begin{minipage}{.028\textwidth}
      \includegraphics[width=0.9\linewidth]{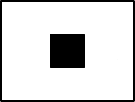}
    \end{minipage}        & 0.0057 & -0.312    & -0.0170  \\
Khan et al. [2022]    &    \begin{minipage}{.028\textwidth}
      \includegraphics[width=0.9\linewidth]{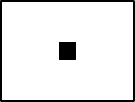}
    \end{minipage}       & 0.0014 & -0.315    & -0.0165  \\
Khan et al. [2022]    &   \begin{minipage}{.028\textwidth}
      \includegraphics[width=0.9\linewidth]{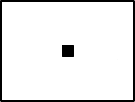}
    \end{minipage}        & 0.0007 & -0.306   & -0.0166  \\
Lorite-D\'iez et al. [2020]  &    \begin{minipage}{.028\textwidth}
      \includegraphics[width=0.9\linewidth]{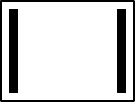}
    \end{minipage}       & 0.035  & -0.181     & -0.0036  \\
Lorite-D\'iez et al. [2020]  &    \begin{minipage}{.028\textwidth}
      \includegraphics[width=0.9\linewidth]{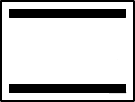}
    \end{minipage}       & 0.035  & -0.291     & -0.0093  \\ \hline
\end{tabular}
\caption{Comparison among the blowing devices studied in the present study, \citet{Khan2022} and \citet{Lorite20}. We include the ratio between blowing slot and the body base areas, $S_b/hw$, the slope of the base drag reduction, $\partial \Delta C_B/\partial(C_q(hw/S_b)^{1/2})$, and the maximum base drag reduction achieved by each experiment, $\Delta C_{B,max}$.}
\label{tab:Discussion}
\end{table}

\section{Conclusions}\label{sec:Conclusions}

Four different blowing slots, each representing $10\%$ of the body base area, referred to as square (S), vertical (V), horizontal (H), and cross (C) configurations, were tested and compared in the present experimental study to evaluate their efficiency in reducing the drag of a square-back Ahmed body by modifying its near-wake characteristics (see Fig.~\ref{fig:PIV_Gy_Gz_Ref}). The wind tunnel experiments, performed at $Re \simeq 65000$, comprised base pressure, aerodynamic forces, and near wake velocity measurements (see Fig.~\ref{fig:setup}). For $C_{q}<C_{q,opt}$, all configurations reveal base pressure recovery (Fig.~\ref{fig:ACx_ACB}), resulting in a drag reduction that is approximately linear with both the injected flow rate and recirculation bubble length ($-\Delta C_{x}\propto C_q$ and $\Delta C_{x}\propto -\Delta L^{*}_{r}$). This behavior is referred to as the mass regime because the jet effectively fills the recirculation bubble. As shown in Fig.~\ref{fig:Discussion}a, the consistent slope across configurations indicates minimal influence of geometry on the wake due to the low jet momentum. In terms of wake asymmetry, for \(0 < C_q < C_{q,opt}\), the blowing configuration has little impact on the RSB mode amplitude (Fig.~\ref{fig:Contour_GyGz}).

The differences between the configurations become apparent at the optimal blowing flow rate, \(C_{q,opt}\), where the jet momentum becomes significant, signaling the end of the mass regime (see Figs.~\ref{fig:ACx_ACB} and ~\ref{fig:Discussion}a). At this critical point, blowing begins to influence three key factors that directly affect drag: bubble elongation (see Fig.~\ref{fig:HORPIV_AVG}–\ref{fig:DeltaFlow_VER}), flow exchanges within the recirculation bubble (see Fig. ~\ref{fig:Backflow}–\ref{fig:VERPIV_RRI}), and wake symmetry (see Fig.~\ref{fig:Contour_GyGz}). At \(C_{q,opt}\), the jet fills the recirculation bubble, displacing the vorticity cores downstream and extending the bubble to its maximum length. This elongation is complemented by the reduced backflow within the recirculation bubble (see Fig. ~\ref{fig:Backflow}), which limits the flux exchange with unperturbed flow (see Figs.~\ref{fig:HORPIV_RRI},~\ref{fig:VERPIV_RRI}). This reduces the turbulence production in the shear layers  (see Figs.~\ref{fig:HORPIV_k} and ~\ref{fig:VERPIV_k}), which is a phenomenon associated with drag reduction. 
Slots with jets acting horizontally (S and H) provide greater drag reduction due to the wider geometry of the model (\(w > h\)) and the greater influence of the horizontal plane on the overall drag compared with the vertical plane. The S configuration is shown to reduce drag by 5$\%$ $C_x$ (12$\%$ in $C_B$) due to an efficient injection of mass and momentum into the recirculation region, resulting in an elongated and less turbulent near wake. The distance between the blowing slot and the shear layers also plays a role. Central slots reduce the drag more effectively than perimetric slots, as supported by \citet{Lorite20, Veerasamy2022,Khan2022}. In the present study, drag reduction increases as the blowing slit approaches the center of the base. This relationship, along with the effect of the body aspect ratio, results in the hierarchy  $\Delta C_{x,max}^S>\Delta C_{x,max}^H$ and $\Delta C_{x,max}^C>\Delta C_{x,max}^V$ between the blowing configurations. Although the drag reduction exceeds that reported by \citet{Khan2022,Khan2024} for smaller central slots, the consistent critical $C_{q,opt}(hw/S_b)^{1/2}$ suggests the existence of a limit driven by the momentum of the jet. 

Once the critical blowing flow rate is reached, the momentum of the blowing jet can alter the spatial characteristics and asymmetry of the near-wake. The momentum regime is governed by the blowing geometry, and it is seen to affect the Reflectional Symmetry Breaking (RSB) mode, typically present in 3D wakes. Similar to pulsed jets \citep{Haffner2021} and perimetric blowing \cite{Lorite20}, configurations that enhance wake asymmetry during the momentum regime  (as V) lead to a dramatic increase in drag with increasing blowing flow rate (Figs. \ref{fig:Contour_GyGz}, \ref{fig:Discussion}c, d) while blowing configurations that reduce the wake asymmetry (H) or symmetrize the flow (S)  lead to lower drag rate of increase at higher flow rates ($C_{q}>C_{q,opt}$) (see Figs. \ref{fig:Contour_GyGz}, \ref{fig:Discussion}c, d). In contrast, the V slot substantially increases the horizontal pressure gradient, which is highly detrimental in terms of drag. Finally, the C slot exhibits a hybrid behavior between H and V. No significant variations in vertical asymmetry due to proximity to the ground were observed in any of the blowing systems evaluated. The suppression of wake asymmetry by configuration S is observed in the momentum regime, as documented in \citet{Veerasamy2022} for a sweeping jet actuator with a blowing area of 5$\%$ of the Ahmed body base. Conversely, smaller central blowing slots achieve a minimal wake asymmetry at lower blowing flow rates \cite{Khan2022, Khan2024}. A comparison of the hierarchy between the presently tested configurations and previous blowing devices analyzed in the literature could help in designing efficient and simple blowing devices acting on 3D blunt bodies.

%%%%%%%%%%%%%%%%%%%%%%%%%%%%%%%%%%%%%%%%%%%%%%%%%%%%%%%%%
%%%%%%%%%%%%%%%%%%%%%%%%%%%%%%%%%%%%%%%%%%%%%%%%%%%%%%%%%
\begin{acknowledgments}

This work is a result of Projects TED2021-131805B-C21 and TED2021-131805B-C22, financed by the Spanish MCIN/AEI/10.13039/501100011033/ and the European Union NextGenerationEU/PRTR. M.L.D. also acknowledges the grant   RYC2023-044496-I financed by MICIU/AEI /10.13039/501100011033 and FSE+. 

\end{acknowledgments}

\section*{References}
\bibliography{PRF_Blowing_2025}
%\begin{thebibliography}

\end{document}